\documentclass[acmsmall]{acmart}
\AtBeginDocument{%
  }

\setcopyright{none}

\usepackage{subcaption}
\usepackage{algorithm2e}
\usepackage{algorithmic}

\begin{document}

\title{Subitizing-Inspired Large Language Models for Floorplanning}

\author{Shao-Chien Lu}
\email{s1111534@mail.yzu.edu.tw}
\orcid{0009-0008-8373-5411}
\author{Chen-Chen Yeh}
\email{s1111511@mail.yzu.edu.tw}
\orcid{0009-0007-1180-4994}
\author{Hui-Lin Cho}
\email{s1112003@mail.yzu.edu.tw}
\orcid{0009-0003-3819-0364}
\author{Yu-Cheng Lin}
\email{linyu@saturn.yzu.edu.tw}
\orcid{0000-0003-2967-0764}
\author{Rung-Bin Lin}
\email{csrlin@saturn.yzu.edu.tw}
\affiliation{%
  \institution{Department of Computer Science and Engineering, Yuan Ze University}
  \city{Taoyuan}
  \country{Taiwan}
}

\renewcommand{\shortauthors}{S.C. Lu et al.}

\begin{abstract}
  We present a novel approach to solving the floorplanning problem by leveraging fine-tuned Large Language Models (LLMs). Inspired by subitizing—the human ability to instantly and accurately count small numbers of items at a glance—we hypothesize that LLMs can similarly address floorplanning challenges swiftly and accurately. We propose an efficient representation of the floorplanning problem and introduce a method for generating high-quality datasets tailored for model fine-tuning. We fine-tune LLMs on datasets with a specified number of modules to test whether LLMs can emulate the human ability to quickly count and arrange items. Our experimental results demonstrate that fine-tuned LLMs, particularly GPT4o-mini, achieve high success and optimal rates while attaining relatively low average dead space. These findings underscore the potential of LLMs as promising solutions for complex optimization tasks in VLSI design.
\end{abstract}




\maketitle

\section{Introduction}
As semiconductor manufacturing processes advance, chip designs increasingly incorporate various Intellectual Properties (IPs) to expedite development cycles. While integrating IPs accelerates the design process, it concurrently amplifies the complexity of chip layout allocation. Floorplanning, a pivotal stage in the Electronic Design Automation (EDA) workflow, entails the strategic placement of functional modules on a chip during the initial design phase. Effective floorplanning must navigate multiple constraints, including minimizing dead space and wire length, managing thermal dissipation, and optimizing the utilization of chip real estate. Additionally, the ability to rotate modules or adjust their width-to-height ratios can enhance their fitting within the designated chip area, further complicating the layout process.

Over the years, several representations for floorplanning results have been introduced, such as sequence pairs \cite{480159}, normalized Polish expressions \cite{wong1986new}, slicing trees \cite{otten1982automatic}, and B*-trees \cite{chang2000b}. Due to the NP-hard nature of the floorplanning problem, traditional approaches have predominantly employed optimization techniques like Simulated Annealing (SA) and Integer Linear Programming (ILP). These methods have demonstrated varying degrees of success in navigating the vast solution space inherent to floorplanning.

\subsection{Traditional Floorplanning Methods}
Simulated Annealing (SA)-based floorplanning has been extensively explored in early research. For instance, \cite{wong1986new} utilized normalized Polish expressions in conjunction with SA to design efficient floorplans. Building upon this, \cite{cheng2005floorplanning} extended SA methodologies to accommodate three-dimensional floorplanning. Further advancements were made by \cite{chen2006modern}, who introduced a fast three-stage SA algorithm based on B*-tree representation, achieving high success rates with reduced dead space. Similarly, \cite{chen2010hybrid} presented a hybrid SA approach that incorporated a novel greedy strategy alongside B*-tree representations, enhancing non-slicing floorplanning accuracy.

In parallel, Integer Linear Programming (ILP) approaches have been investigated for floorplanning. \cite{sutanthavibul1991analytical} employed ILP to simultaneously minimize interconnection delays and chip area, addressing critical performance metrics in floorplanning. Additionally, \cite{wu2005reticle} demonstrated the application of ILP in reticle design and wafer dicing processes for multiple project wafers, highlighting its versatility in related semiconductor manufacturing tasks.

\subsection{Machine Learning Approaches}
The advent of machine learning (ML) has opened new avenues for addressing floorplanning challenges. Researchers have increasingly explored ML-based solutions to complement or replace traditional optimization techniques. For example, \cite{xu2021goodfloorplan} introduced GoodFloorplan, which leverages reinforcement learning to minimize both area and wire length, yielding impressive results on MCNC and GSRC benchmarks. Building on this, \cite{yu2024deep} adopted deep reinforcement learning strategies based on sequence pairs to achieve superior floorplanning solutions. Furthermore, \cite{liu2023hybrid} integrated hybrid reinforcement learning with genetic algorithms (GA), demonstrating significant reductions in wire length and area in the resultant floorplans.

\subsection{Large Language Models in Floorplanning}
Large Language Models (LLMs) have revolutionized various domains by automating complex tasks and enhancing decision-making processes. Despite their transformative impact, the application of LLMs to floorplanning remains relatively unexplored, primarily due to challenges related to data representation and the scarcity of large-scale, high-quality datasets. Inspired by the concept of subitizing, as first introduced by \cite{kaufman1949discrimination}, which describes the human ability to instantly and accurately recognize the number of items in a small set, we hypothesize that LLMs can similarly provide rapid and accurate solutions to floorplanning problems after exposure to a sufficient number of examples.

Supporting this hypothesis, \cite{trick1994small} provided experimental evidence indicating that human reaction times for recognizing fewer than four items range between 40 to 100 milliseconds per item. However, for larger quantities, reaction times escalate to 250 to 350 milliseconds, suggesting a shift in cognitive processing strategies. They proposed three theories of enumeration: density-based, pattern-based, and working memory explanations. Drawing an analogy, we posit that LLMs can efficiently recognize and generate optimal solutions for small-scale floorplanning tasks by leveraging pattern recognition and memory-based strategies akin to human cognitive processes.

\subsection{Contributions}
We harness the capabilities of fine-tuned Large Language Models (LLMs) to address the floorplanning problem. Our primary contributions are as follows:

\begin{itemize}
  \item We define a concise and effective representation for the floorplanning problem, facilitating efficient processing by LLMs.
  \item We propose an innovative method for generating high-quality datasets tailored for the fine-tuning of LLMs, ensuring robustness and scalability.
  \item We present the first successful application of fine-tuned LLMs to solve floorplanning problems, demonstrating their potential to outperform traditional and contemporary ML-based methods.
\end{itemize}

The paper is organized as follows: Section~\ref{sec:problem_description} provides a detailed description of the floorplanning problem, including the calculation of dead space and the iterative process for optimizing module placements. Section~\ref{sec:methodology} outlines our proposed methodology, which comprises two primary stages: fine-tuning and inference. Section~\ref{sec:experimental_results} presents the experimental results of our approach, comparing two methodologies—local fine-tuning with Unsloth and leveraging the OpenAI API to fine-tune GPT4o-mini. Finally, Section~\ref{sec:conclusion} concludes the paper and outlines future research directions.

\section{Problem Description}
\label{sec:problem_description}
\begin{figure}[h]
  \centering
  \includegraphics[width=0.6\textwidth]{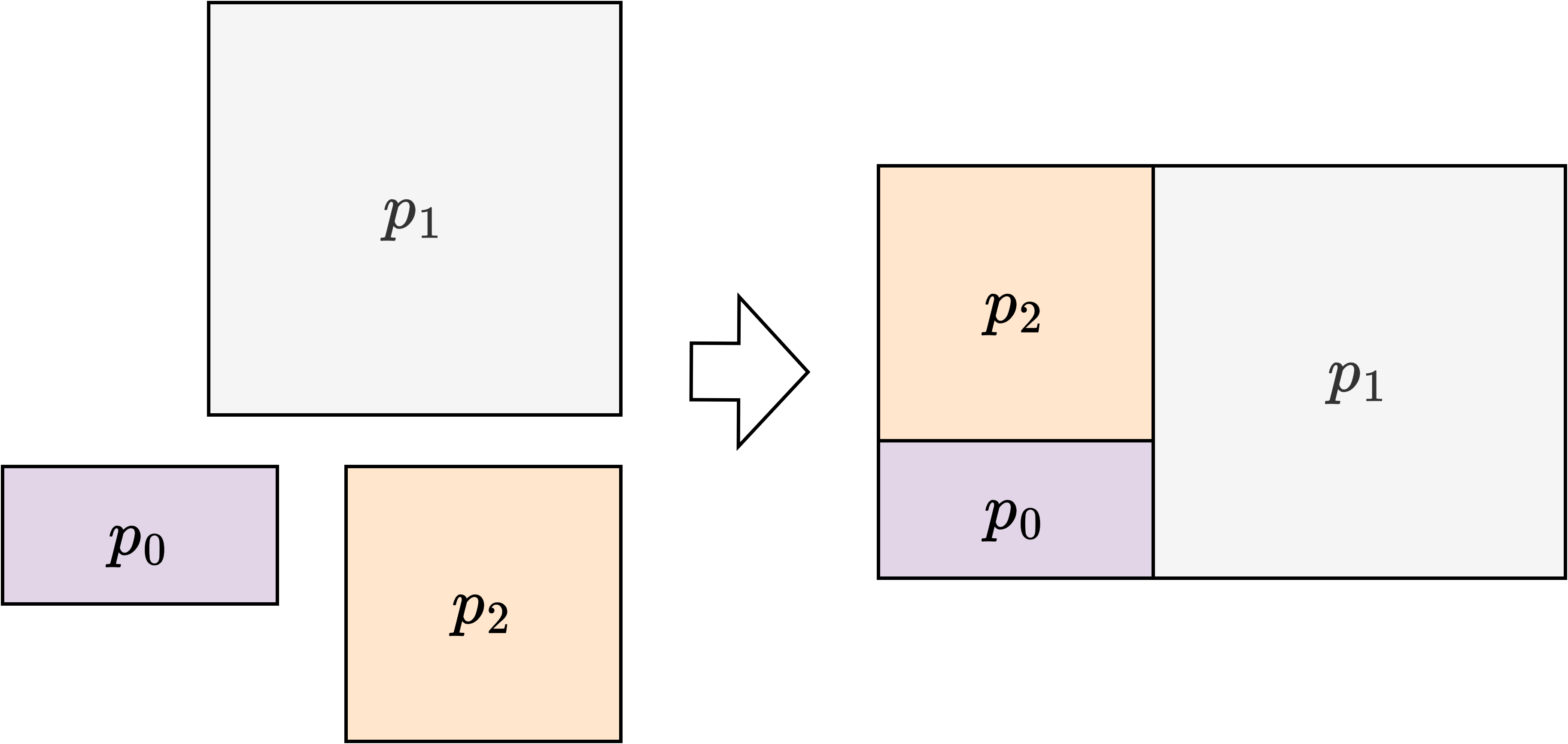}
  \caption{Floorplanning problem with three modules}
  \Description{We use three modules, $p_0$ to $p_2$, to illustrate the floorplanning problem. After getting three modules, we need to arrange them in a two-dimensional space to minimize dead space. The dead space is the unused area that arises from gaps between adjacent modules. In this figure, the modules are arranged in a way that minimizes the dead space. While there are multiple ways to arrange the modules, the goal is to find the optimal arrangement that minimizes the dead space.}
  \label{fig:floorplan_problem}
\end{figure}

In the initial stages of the Electronic Design Automation (EDA) workflow, floorplanning involves determining the approximate placement of each module on a chip. Given a set of $n$ modules $P = \{p_1, p_2, \ldots, p_n\}$, each with specified dimensions—width $w_i$ and height $h_i$—the objective is to arrange all modules within a two-dimensional space to minimize dead space. In this context, each module possesses a fixed shape, prohibiting rotations or dimensional adjustments to simplify the floorplanning process for LLMs. For example, Figure~\ref{fig:floorplan_problem} illustrates three modules needing to be floorplanned. One possible floorplan is shown in the figure. As the number of modules increases, the problem becomes significantly more complex.

To record floorplanning results, a straightforward approach is to directly store the coordinates of each module. However, this method becomes inefficient for large-scale floorplanning tasks, requiring substantial memory and computational resources. For representing a floorplan, there are two primary categories: non-slicing and slicing floorplans. A slicing floorplan implies that modules are obtained through horizontal or vertical cuts, whereas a non-slicing floorplan captures the relative vertical and horizontal relationships between modules.

\begin{figure}[h]
  \centering
  \includegraphics[width=0.6\textwidth]{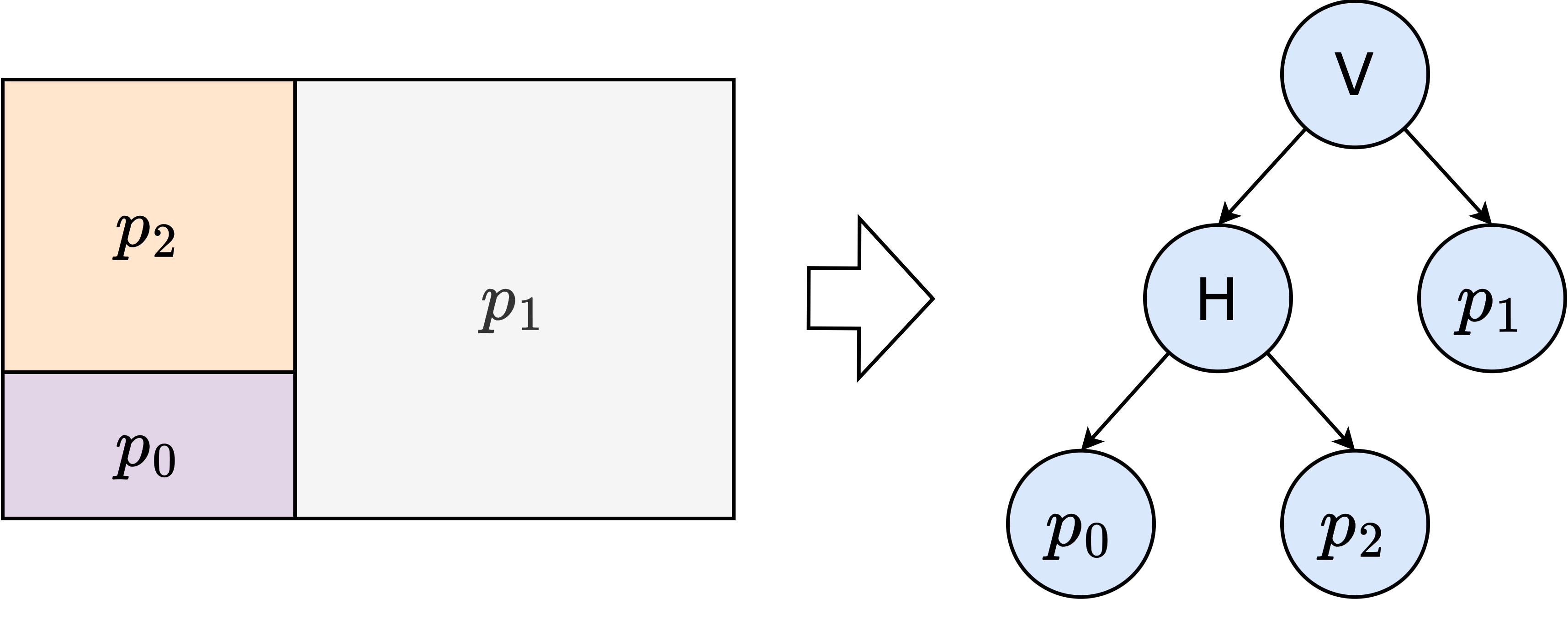}
  \caption{Turning a slicing floorplan into a slicing tree}
  \Description{Given a slicing floorplan, we can convert it into a slicing tree. The slicing tree is a binary tree that represents the floorplan. The tree is constructed by recursively slicing the modules horizontally or vertically. In this example, we first slice vertically to separate $p_1$ from the composite module containing $p_0$ and $p_2$. Then, we slice horizontally to separate $p_0$ and $p_2$. The final tree structure is shown in the figure.}
  \label{fig:slicing_example}
\end{figure}

This paper focuses on slicing floorplan representation, which, although potentially losing some possible floorplans, offers easier generation and storage. Figure~\ref{fig:slicing_example} illustrates a slicing floorplan and its corresponding slicing tree. We first slice vertically to separate $p_1$ from the composite module containing $p_0$ and $p_2$, denoting the root node as "V" (vertical slicing). The left (or bottom) module is assigned as the left child node to maintain consistency. Subsequently, a horizontal slice separates $p_0$ and $p_2$, represented by the "H" in the tree. The final tree structure is shown in the figure.

Dead space is defined as the unused area that arises from gaps between adjacent modules. The calculation of dead space depends on the relative positioning of the modules. Specifically, when two modules are placed side by side along the $x$-axis, the dead space is computed as the product of the width of the module with the smaller height and the absolute difference in their heights. Similarly, when two modules are stacked along the $y$-axis, the dead space is computed as the product of the height of the module with the smaller width and the absolute difference in their widths.

For two adjacent modules \(p_i\) and \(p_j\) with dimensions \((w_i, h_i)\) and \((w_j, h_j)\), respectively, the dead space is defined as follows:

\begin{equation}
  \text{dead\_space}(p_i, p_{i+1}) =
  \begin{cases}
    \text{dead\_space}_{\text{hor}}(p_i, p_{i+1}), & \text{if } p_i \text{ and } p_{i+1} \text{ are horizontally adjacent,} \\[2mm]
    \text{dead\_space}_{\text{ver}}(p_i, p_{i+1}), & \text{if } p_i \text{ and } p_{i+1} \text{ are vertically adjacent.}
  \end{cases}
  \label{eq:dead_space}
\end{equation}

where the \(\text{dead\_space}_{\text{hor}}(p_i, p_{i+1})\) and \(\text{dead\_space}_{\text{ver}}(p_i, p_{i+1})\) functions are defined as follows:

\begin{equation}
  \text{dead\_space}_{\text{hor}}(p_i, p_j) =
  \begin{cases}
    w_i \times |h_i - h_j|, & \text{if } h_i \leq h_j, \\
    w_j \times |h_i - h_j|, & \text{otherwise}.
  \end{cases}
  \label{eq:dead_space_hor}
\end{equation}

\begin{equation}
  \text{dead\_space}_{\text{ver}}(p_i, p_j) =
  \begin{cases}
    h_i \times |w_i - w_j|, & \text{if } w_i \leq w_j, \\
    h_j \times |w_i - w_j|, & \text{otherwise}.
  \end{cases}
  \label{eq:dead_space_ver}
\end{equation}

For scenarios involving more than two modules, the dead space calculation is generalized through an iterative process that selects two modules, calculates their dead space, and merges them into a single composite module. This process continues until all modules are combined into a single composite module. The dimensions of the composite module are updated as follows:

\begin{align}
  \text{if horizontally adjacent:} \quad & w_{\text{new}} = w_i + w_{i+1}, \quad h_{\text{new}} = \max(h_i, h_{i+1}), \\[2mm]
  \text{if vertically adjacent:} \quad   & w_{\text{new}} = \max(w_i, w_{i+1}), \quad h_{\text{new}} = h_i + h_{i+1}.
\end{align}

Algorithm~\ref{alg:dead_space} provides a detailed description of the dead space calculation process for two adjacent modules. The input to the algorithm is a list of nodes \(L\) arranged in post-order. We assume that the input is a valid slicing tree, where each node is either a module node or a slicing node (horizontal or vertical). The algorithm uses a stack to store module nodes. When encountering a slicing node, it pops the two most recent module nodes from the stack, calculates the dead space between them, and merges them into a composite module using the \texttt{merge} function (Algorithm~\ref{alg:merge}). The composite module is then pushed back onto the stack. The process continues until the list is empty, at which point the total dead space is returned.

\begin{figure}[htb]
  \centering
  \begin{minipage}{.55\linewidth}
    \centering
    \begin{algorithm}[H]
      \caption{Dead Space Calculation}
      \label{alg:dead_space}
      \SetKwInOut{Input}{Input}
      \SetKwInOut{Output}{Output}
      \Input{List of nodes \(L\) in post-order traversal}
      \Output{Total dead space \(DS\)}
      \BlankLine
      \SetKwFunction{CalcDS}{CalculateDeadSpace}
      \SetKwProg{Fn}{Function}{:}{}
      \Fn{\CalcDS{\(L\)}}{
        \KwData{Stack \(S\) to store module nodes}
        \KwData{\(DS \leftarrow 0\)}
        \BlankLine
        \While{\(L\) is not empty}{
          \(node \leftarrow L.\text{pop()}\)\;
          \If{\(node\) is a module node}{
            \(S.\text{push}(node)\)\;
          }
          \ElseIf{\(node\) is a slicing node}{
            \If{\(node\) is a horizontal slicing node}{
              \(DS \leftarrow DS + \text{dead\_space}_{\text{hor}}(p_i, p_j)\)\;
              \(p_{\text{new}} \leftarrow \text{merge}(p_i, p_j, \text{horizontal})\)\;
            }
            \Else{
              \(DS \leftarrow DS + \text{dead\_space}_{\text{ver}}(p_i, p_j)\)\;
              \(p_{\text{new}} \leftarrow \text{merge}(p_i, p_j, \text{vertical})\)\;
            }
            \(S.\text{push}(p_{\text{new}})\)\;
          }
        }
        \Return{\(DS\)}\;
      }
    \end{algorithm}
  \end{minipage}
  \hfill
  \begin{minipage}{.4\linewidth}
    \centering
    \begin{algorithm}[H]
      \caption{Merge Function}
      \label{alg:merge}
      \SetKwInOut{Input}{Input}
      \SetKwInOut{Output}{Output}
      \Input{Two modules \(p_i\) and \(p_j\)}
      \Input{Merging direction (horizontal or vertical)}
      \Output{Composite module \(p_{\text{new}}\)}
      \BlankLine
      \SetKwFunction{Merge}{merge}
      \SetKwProg{Fn}{Function}{:}{}
      \Fn{\Merge{\(p_i\), \(p_j\), direction}}{
        \If{direction is horizontal}{
          \(w_{\text{new}} \leftarrow w_i + w_j\)\;
          \(h_{\text{new}} \leftarrow \max(h_i, h_j)\)\;
        }
        \Else{
          \(w_{\text{new}} \leftarrow \max(w_i, w_j)\)\;
          \(h_{\text{new}} \leftarrow h_i + h_j\)\;
        }
        \Return{Module(\(w_{\text{new}}\), \(h_{\text{new}}\))}\;
      }
    \end{algorithm}
  \end{minipage}
  \caption{(Left) Dead space calculation algorithm traversing the slicing tree in post-order and computing the dead space between adjacent modules. (Right) Merge function combining two modules based on the merging direction to create a composite module.}
  \Description{The dead space calculation algorithm (Algorithm~\ref{alg:dead_space}) traverses the slicing tree in post-order and calculates the dead space between adjacent modules. The merge function (Algorithm~\ref{alg:merge}) merges two modules into a composite module based on the merging direction.}
\end{figure}

\begin{figure}[]
  \centering
  \begin{subfigure}[b]{0.23\textwidth}
    \centering
    \includegraphics[width=\textwidth]{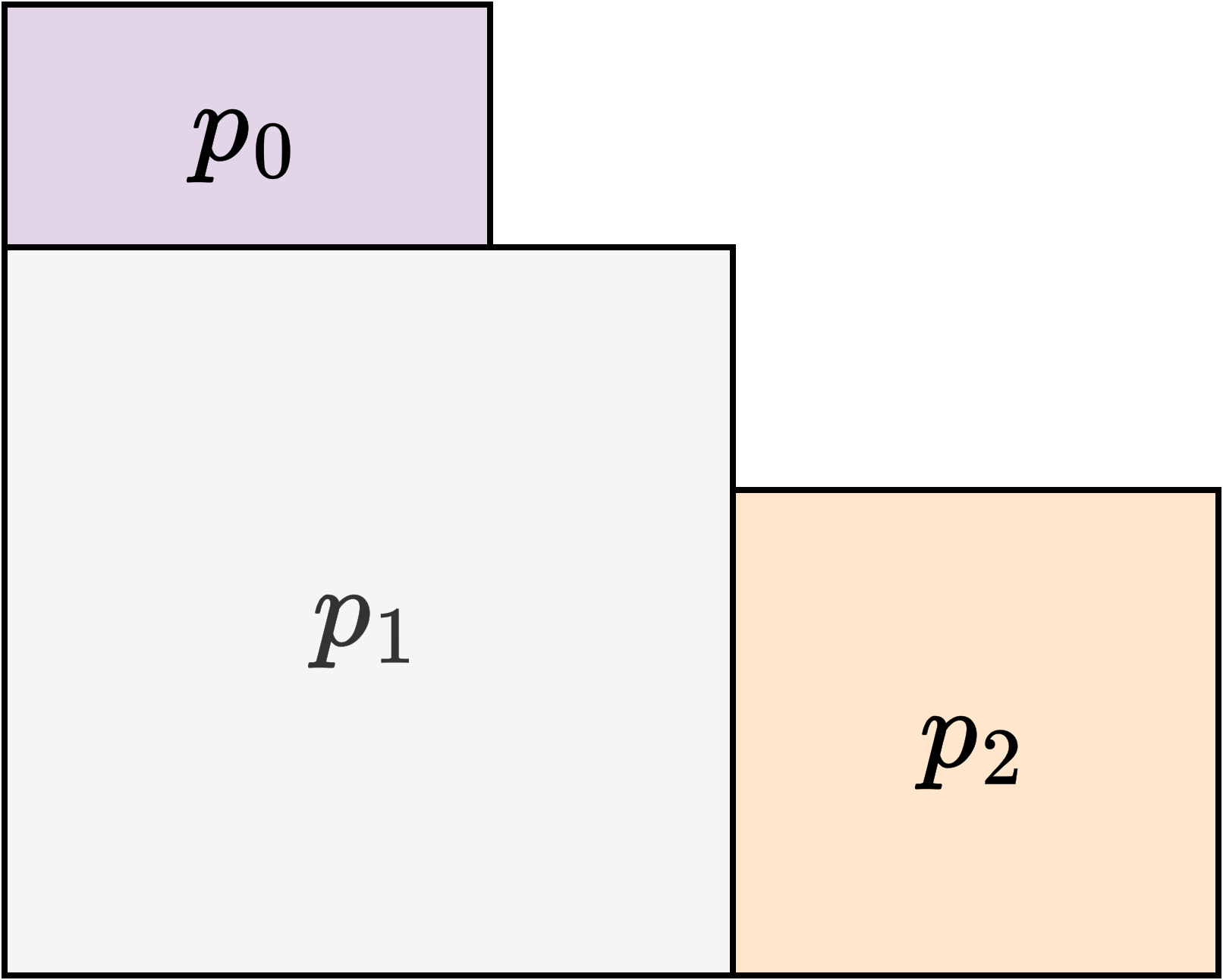}
    \caption{Non-optimal floorplan}
    \Description{This is a non-optimal floorplan with three modules. The $p_0$ is on the top of $p_1$ and $p_2$ is on the right side of $p_1$.}
    \label{fig:deadspace_floorplan}
  \end{subfigure}
  \hfill
  \begin{subfigure}[b]{0.23\textwidth}
    \centering
    \includegraphics[width=\textwidth]{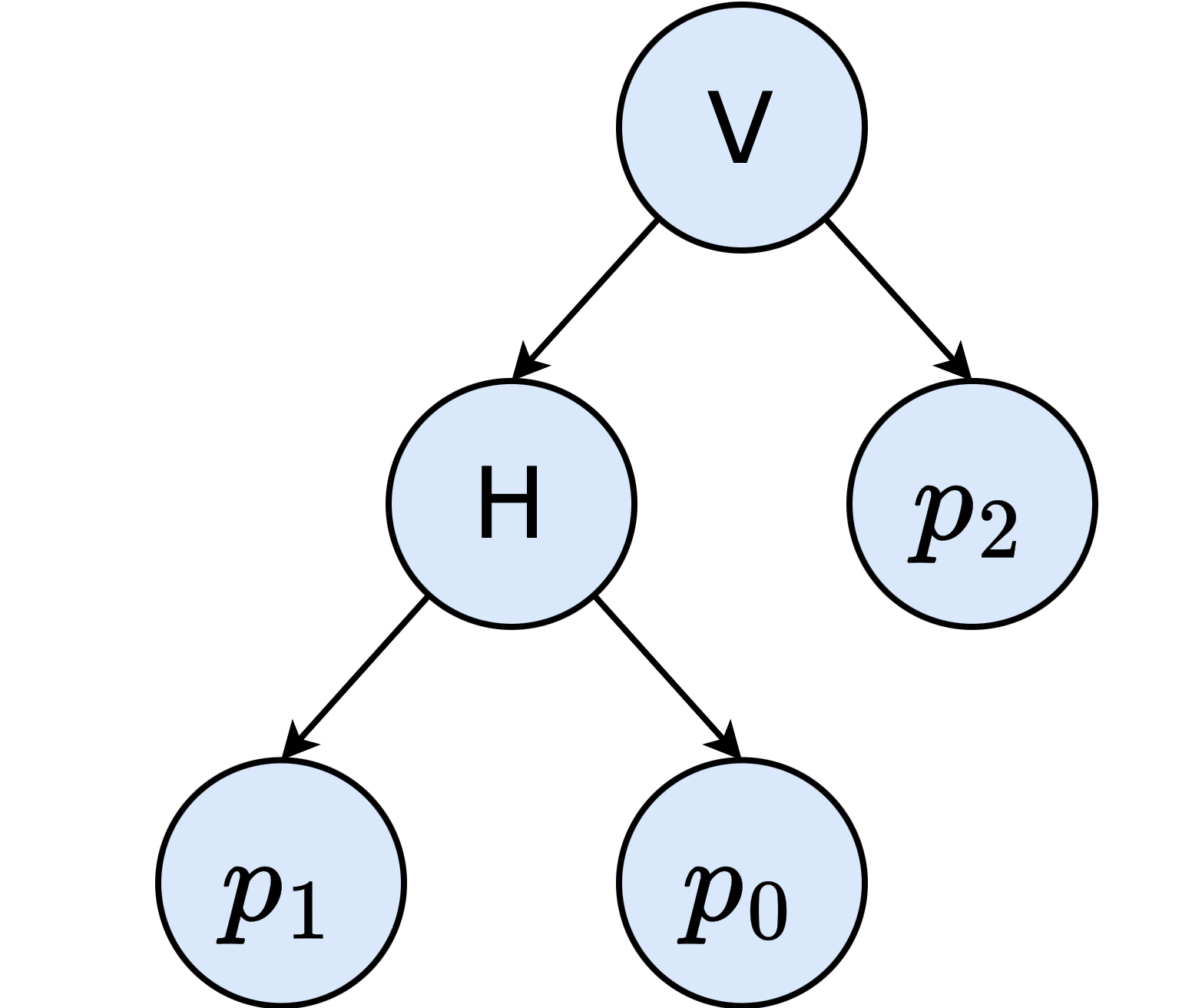}
    \caption{Slicing tree for the floorplan}
    \Description{This is a slicing tree for the floorplan. The root node is a vertical slice, and the left and right child nodes are horizontal slices and $p_2$, respectively. The child nodes of the horizontal slice are $p_1$ and $p_0$.}
    \label{fig:deadspace_tree}
  \end{subfigure}
  \hfill
  \begin{subfigure}[b]{0.255\textwidth}
    \centering
    \includegraphics[width=\textwidth]{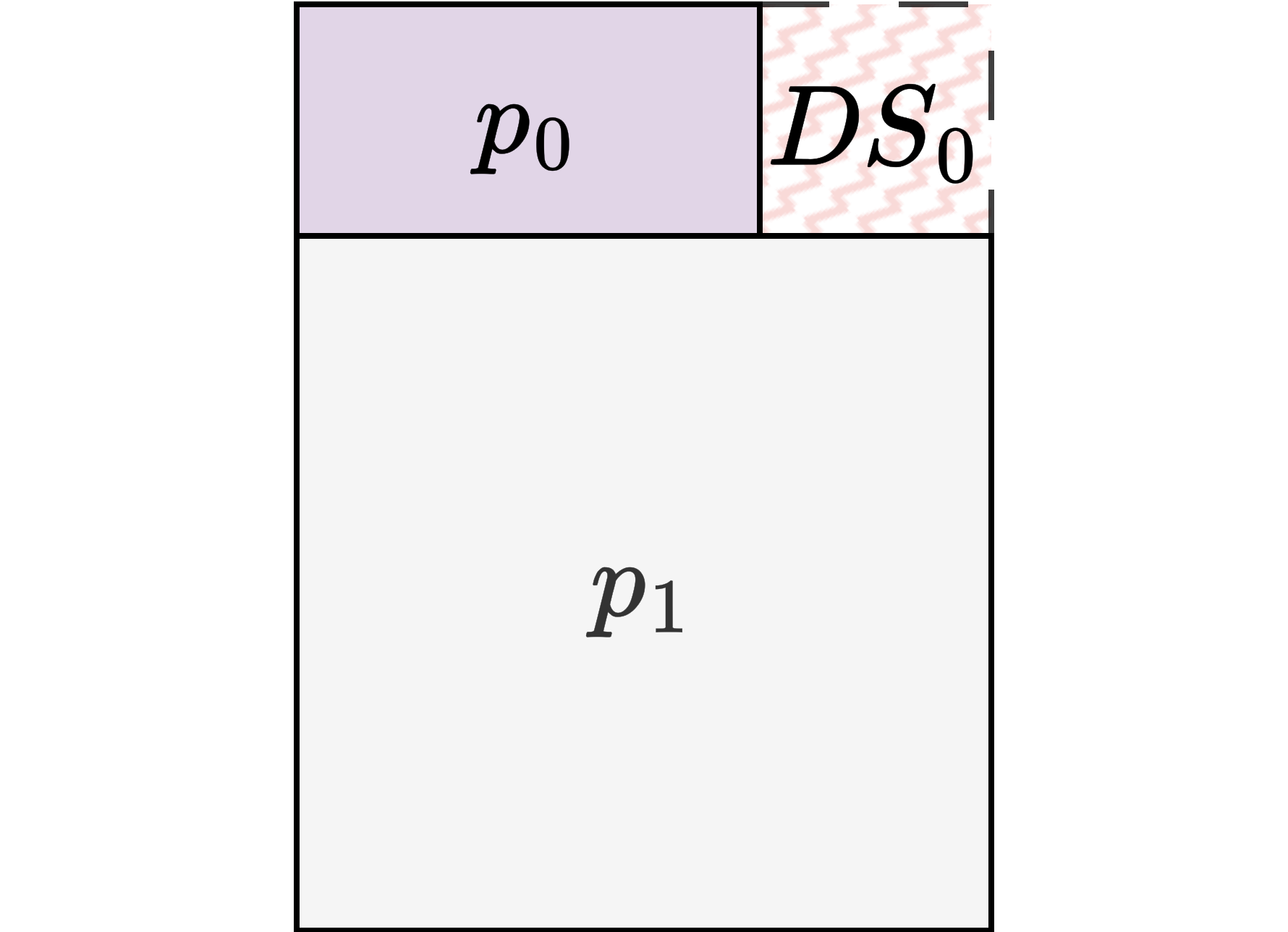}
    \caption{Dead space for horizontal adjacency}
    \Description{The is the dead space calculation for horizontal adjacency since we using the post-order traversal to calculate the dead space. We use $DS_0$ to represent the dead space between $p_0$.}
    \label{fig:deadspace_ds_0}
  \end{subfigure}
  \hfill
  \begin{subfigure}[b]{0.23\textwidth}
    \centering
    \includegraphics[width=\textwidth]{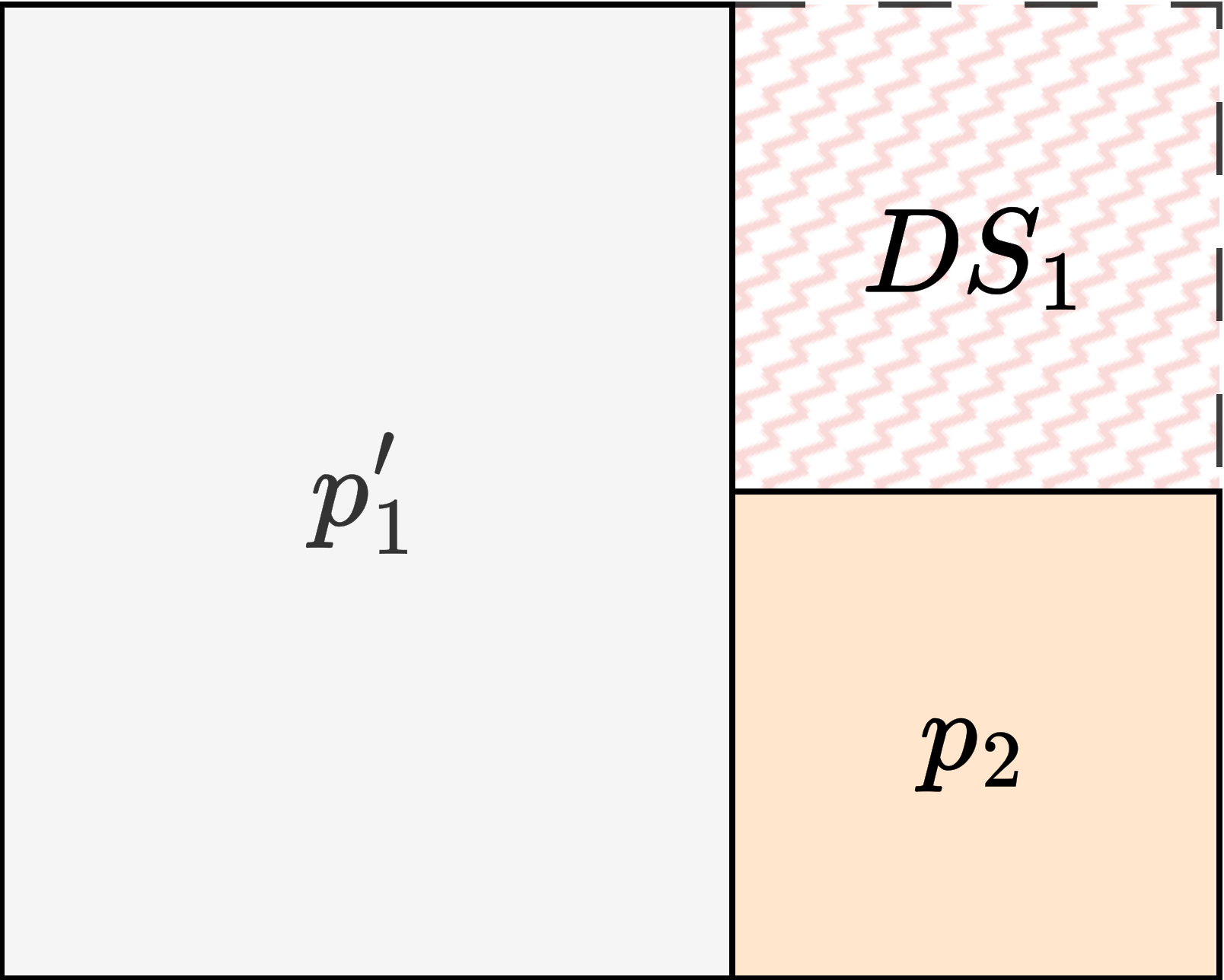}
    \caption{Dead space for vertical adjacency}
    \Description{The is the dead space calculation for vertical adjacency since we using the post-order traversal to calculate the dead space. We use $DS_1$ to represent the dead space between $p_1$ and $p_2$.}
    \label{fig:deadspace_ds_1}
  \end{subfigure}
  \caption{Dead space calculation for floorplanning (DS: Dead Space)}
  \Description{An example of dead space calculation for a non-optimal floorplan step by step.}
  \label{fig:dead_space}
\end{figure}

Figure~\ref{fig:dead_space} illustrates examples of dead space calculations for horizontally and vertically adjacent modules. Figure~\ref{fig:deadspace_floorplan} shows a non-optimal floorplan, and Figure~\ref{fig:deadspace_tree} depicts the corresponding slicing tree. We can traverse the tree in post-order, calculate the dead space between two adjacent modules, merge them into a composite module, and update the dimensions. Figures~\ref{fig:deadspace_ds_0} and~\ref{fig:deadspace_ds_1} show the dead space calculation for horizontal and vertical adjacency, respectively. In this case, the total dead space is computed as $DS_0 + DS_1$. The optimal floorplan is defined as the one with the minimum dead space.

\section{Methodology}
\label{sec:methodology}
\begin{figure}[h]
  \centering
  \begin{subfigure}[b]{0.23\textwidth}
    \centering
    \includegraphics[width=\textwidth]{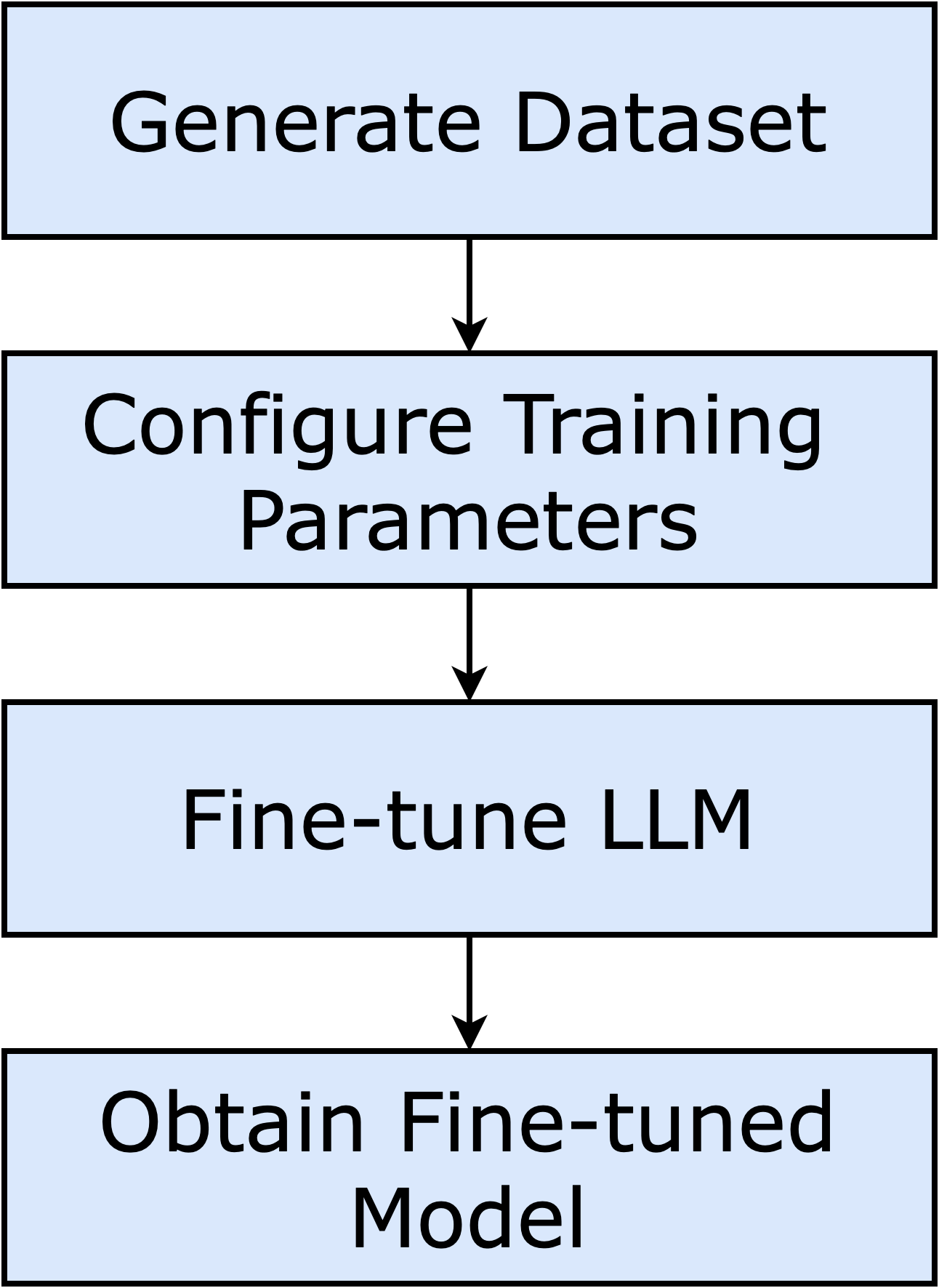}
    \caption{Fine-tuning stage}
    \Description{Flowchart of the fine-tuning stage. We will first generate dataset, configure the training parameters. Then start to fine-tune the model.}
    \label{fig:finetune}
  \end{subfigure}
  \hfill
  \begin{subfigure}[b]{0.75\textwidth}
    \centering
    \includegraphics[width=\textwidth]{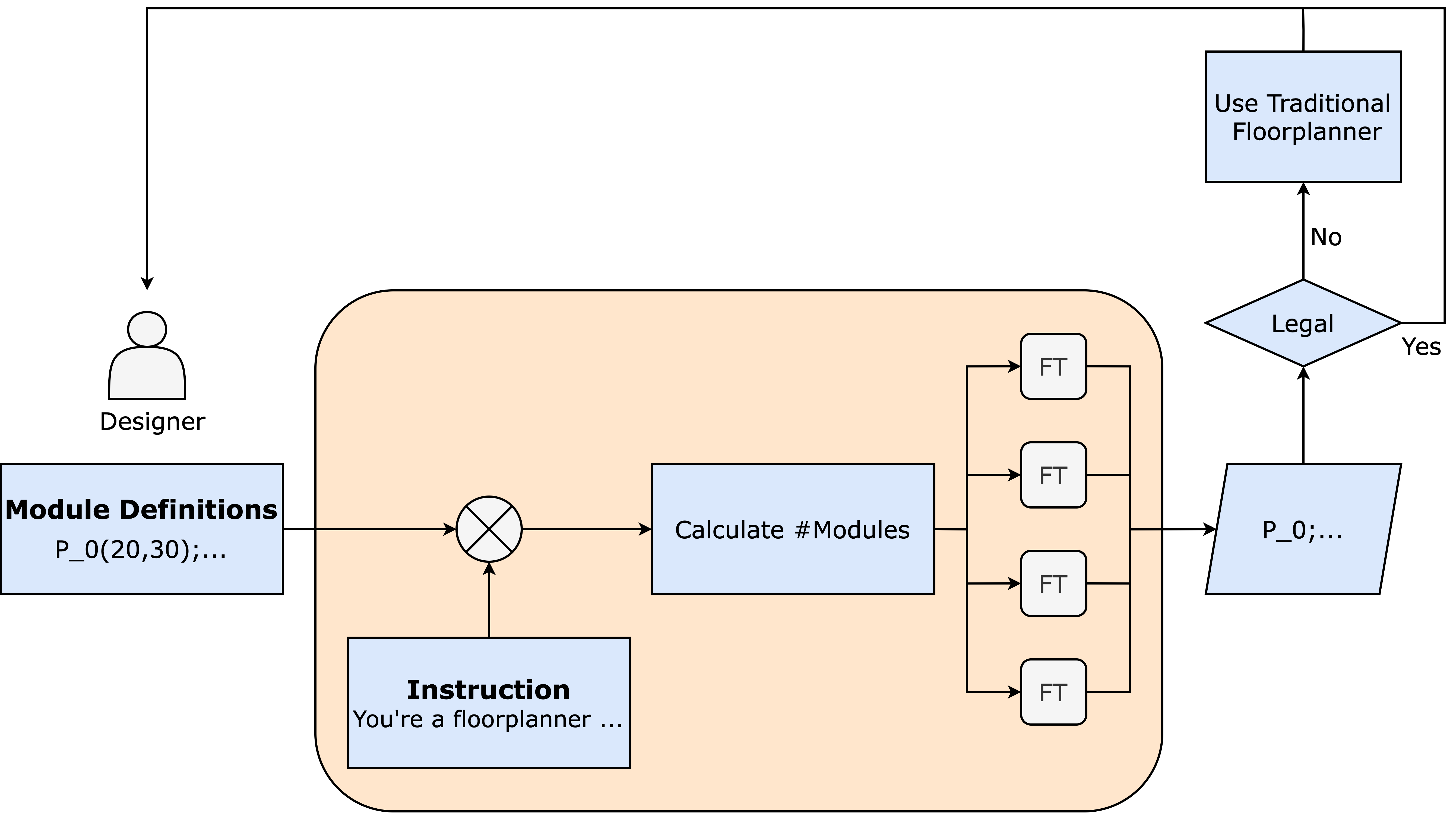}
    \caption{Inference stage (FT: Fine-tuned model)}
    \Description{Flowchart of the inference stage. We will first input the modules definitions to the fine-tuned model. Then the model will generate the slicing tree. Finally, we will integrate the slicing tree with a traditional floorplanner to get the final floorplan.}
    \label{fig:infer}
  \end{subfigure}
  \caption{Overview of our workflow}
  \Description{Overview of the two main stages: fine-tuning and inference}
  \label{fig:workflow}
\end{figure}

Figure~\ref{fig:workflow} illustrates an overview of our proposed workflow, which comprises two primary stages: fine-tuning and inference.

\subsection{Fine-tuning Stage}
\label{subsec:finetuning}

The fine-tuning stage, depicted in Figure~\ref{fig:finetune}, involves generating a comprehensive dataset and subsequently fine-tuning a Large Language Model (LLM) to predict optimal floorplanning solutions.

\begin{figure}[h]
  \centering
  \begin{subfigure}[b]{0.23\textwidth}
    \centering
    \includegraphics[width=\textwidth]{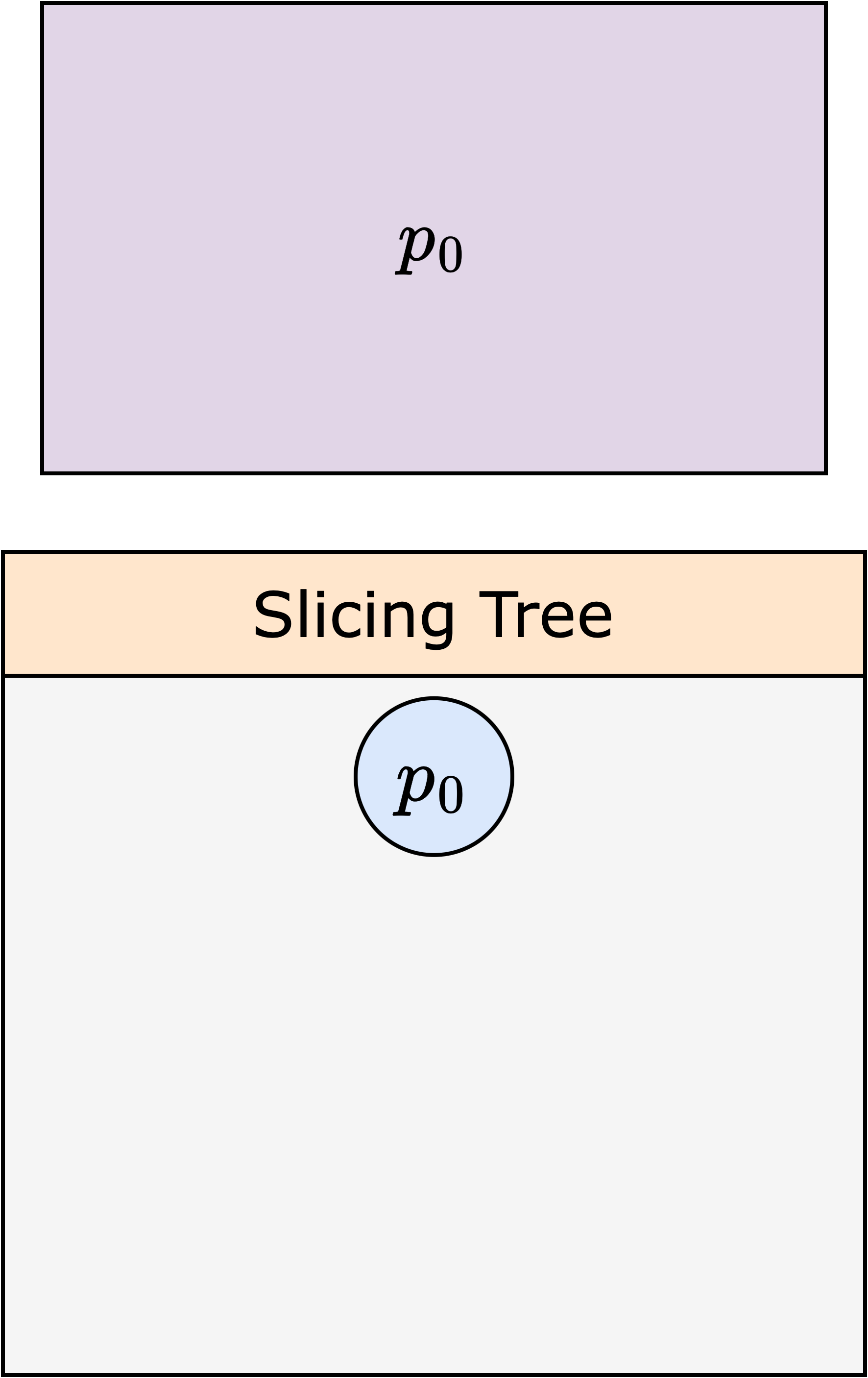}
    \caption{Rectangle Generation}
    \Description{The initial rectangle generation process. We generate a rectangle representing the chip boundary with random width and height.}
    \label{fig:slice_1}
  \end{subfigure}
  \begin{subfigure}[b]{0.23\textwidth}
    \centering
    \includegraphics[width=\textwidth]{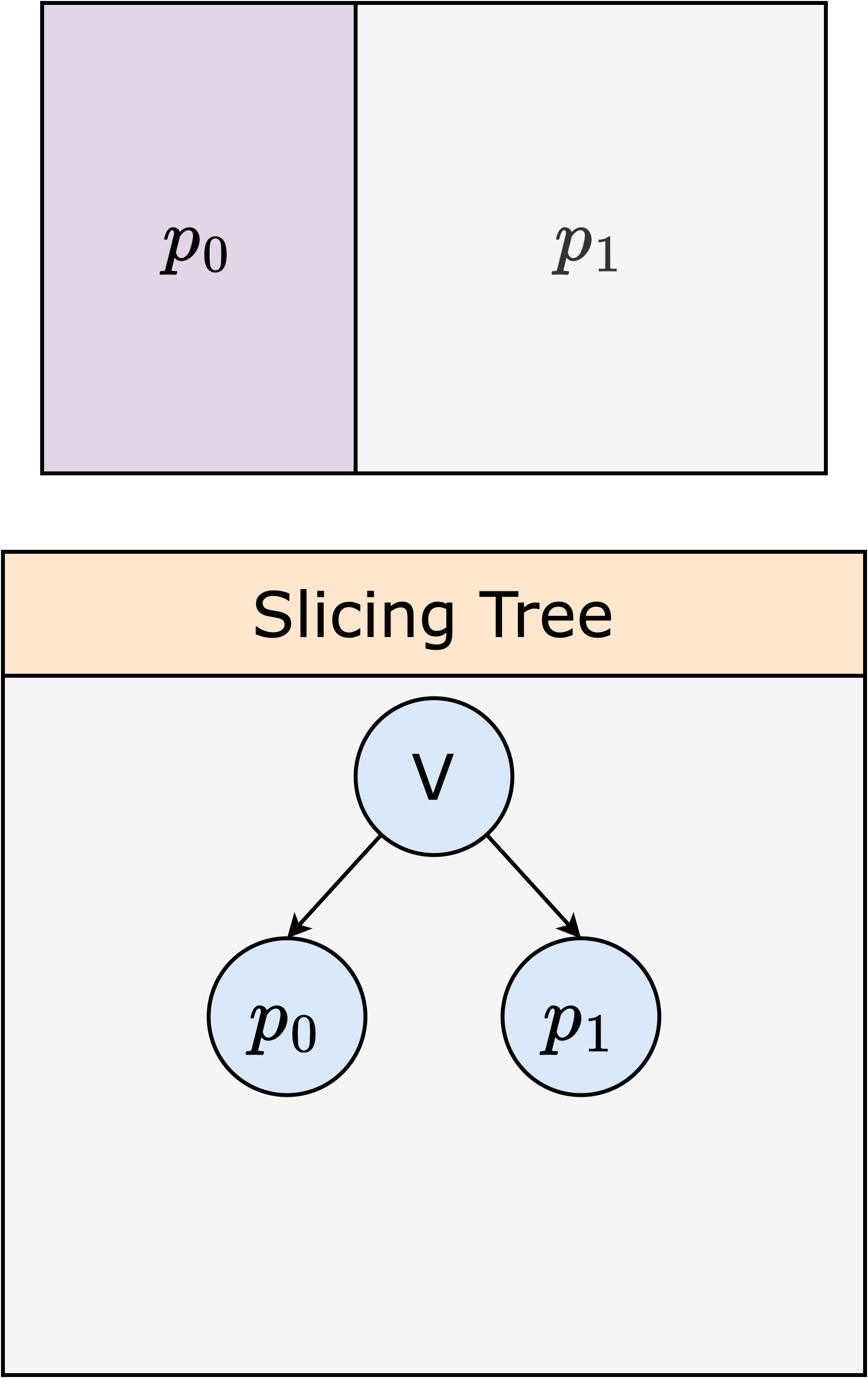}
    \caption{Vertical Slicing}
    \Description{Slice the rectangle vertically to separate two modules. The left module is designated as the left child node in the slicing tree.}
    \label{fig:slice_2}
  \end{subfigure}
  \begin{subfigure}[b]{0.23\textwidth}
    \centering
    \includegraphics[width=\textwidth]{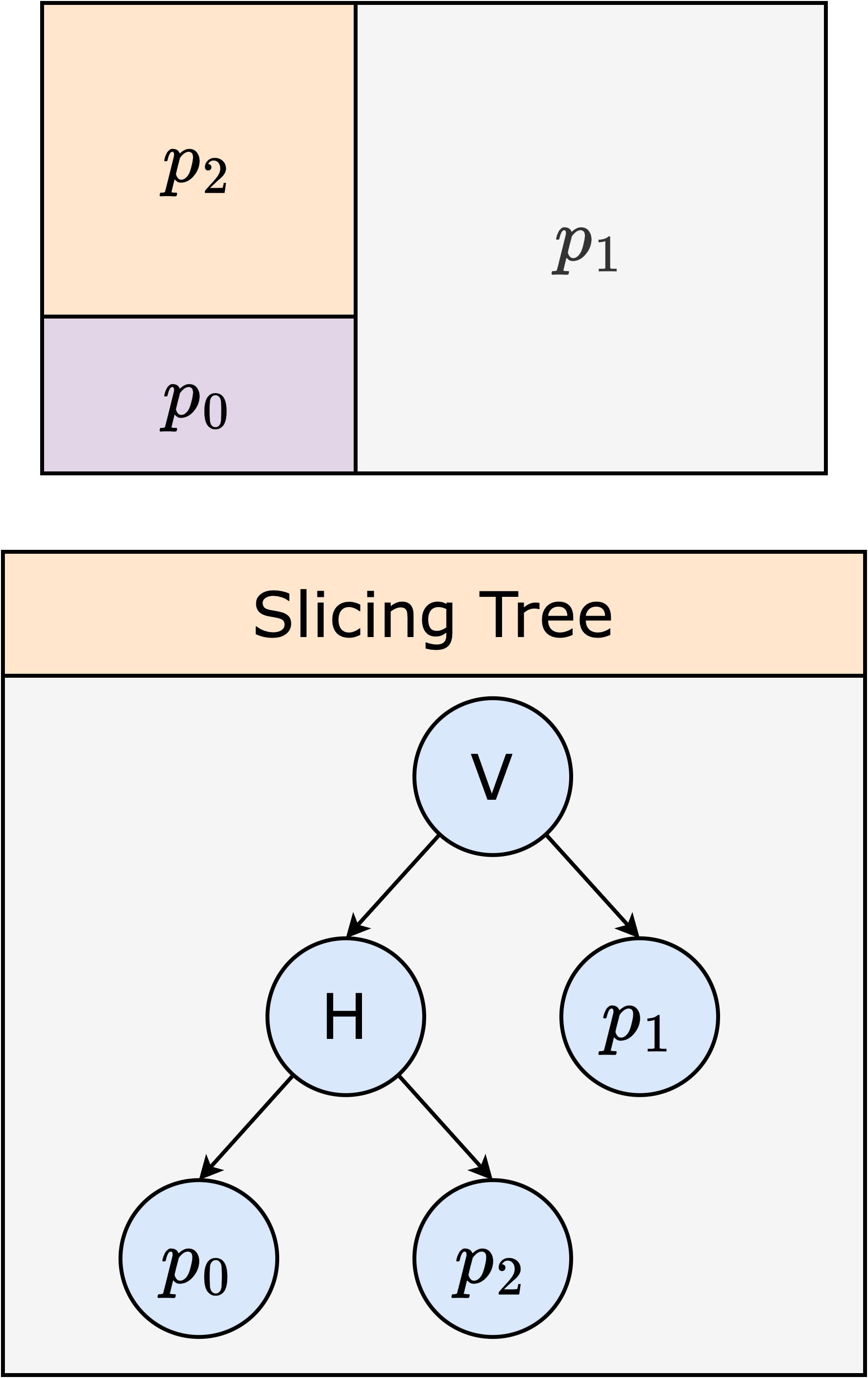}
    \caption{Horizontal Slicing}
    \Description{Slice the rectangle horizontally to separate two modules. The left module is designated as the left child node in the slicing tree.}
    \label{fig:slice_3}
  \end{subfigure}
  \begin{subfigure}[b]{0.23\textwidth}
    \centering
    \includegraphics[width=\textwidth]{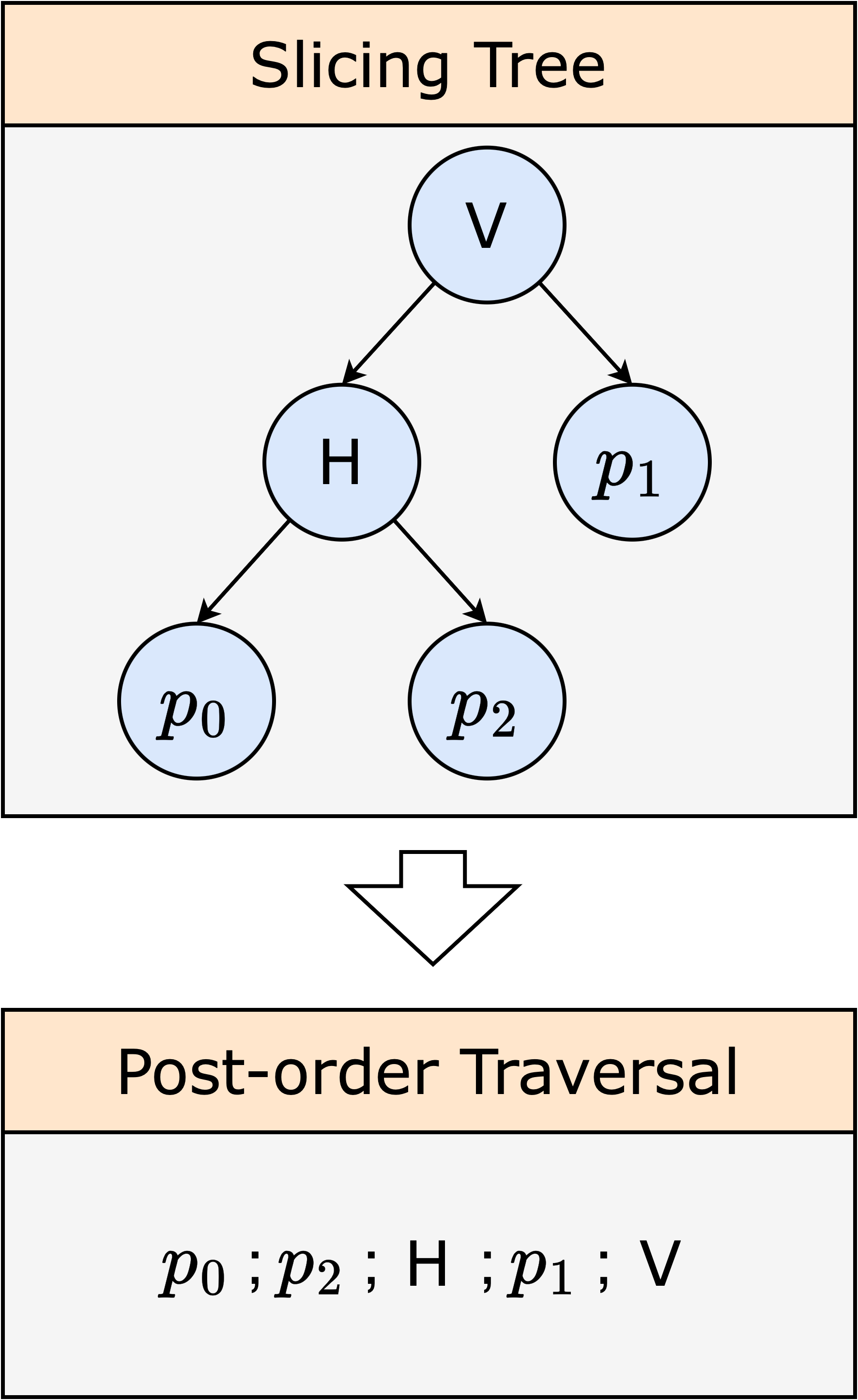}
    \caption{Tree Encoding}
    \Description{Traverse the slicing tree in post-order to encode the tree. Each module is labeled as \(p_i\), and slicing operations are denoted by 'H' or 'V'.}
    \label{fig:encoding}
  \end{subfigure}
  \caption{Illustration of the recursive slicing and tree encoding process.}
  \Description{A diagram showing rectangle generation, slicing operations, and tree encoding.}
  \label{fig:slicing}
\end{figure}

\textbf{Dataset Generation:}
The scarcity of extensive floorplanning datasets has historically limited the effectiveness of machine learning approaches in this domain. While one might consider using an existing floorplanner to generate large datasets, this approach is time-consuming and relies on obtaining circuit data (e.g., from the MCNC benchmarks), which is not always feasible. Moreover, even real data generated by floorplanners may not be optimal. To overcome these challenges, we introduce a novel data generation technique based on recursive slicing and tree encoding. The process, illustrated in Figure~\ref{fig:slicing}, comprises the following steps:
\begin{enumerate}
  \item \textbf{Initial Rectangle Generation:} Generate a rectangle representing the chip boundary with random width and height.
  \item \textbf{Recursive Slicing:}
        \begin{itemize}
          \item Randomly select a rectangle (module) and slice it either horizontally or vertically.
          \item Each slicing operation yields two adjacent modules; the slicing direction determines whether the width or height remains constant.
          \item Designate the left (or bottom) module as the left child node in the slicing tree.
          \item Continue the recursive slicing until a complete slicing tree is formed.
        \end{itemize}
  \item \textbf{Tree Encoding:} Encode the resultant slicing tree using post-order traversal. Each module is labeled as \(p_i\), and slicing operations are denoted by 'H' (horizontal) or 'V' (vertical), separated by semicolons.
\end{enumerate}

We choose a slicing tree representation because it enables easier dataset generation compared to alternative methods. In particular, using post-order traversal (instead of pre-order) avoids the generation of long sequences of consecutive "H" or "V" operations, which can lead to errors in LLM outputs. For example, a pre-order traversal might yield "V;H;\(p_0\);\(p_2\);\(p_1\)," which becomes increasingly problematic as dataset size grows. This encoding effectively captures the hierarchical structure of the floorplan, enabling the creation of large-scale datasets necessary for robust LLM fine-tuning.

The fine-tuning process involves training the LLM on datasets tailored to specific module counts (16 and 24 modules) to validate our hypothesis that LLMs can perform floorplanning in a manner akin to human subitizing. The input to the LLM consists of module names along with their respective widths and heights, combined with background instructions. The desired output is a slicing tree in post-order traversal that represents the optimal floorplan.

\subsection{Inference Stage}
\label{subsec:inference}

In the inference stage, as illustrated in Figure~\ref{fig:infer}, the fine-tuned LLM receives a set of modules as input and generates a corresponding floorplanning solution. However, due to inherent limitations in language models, the LLM occasionally produces not optimal results or even illegal slicing trees. Consequently, the LLM's output cannot be used directly; instead, it should be integrated with a traditional floorplanner to provide guidance, rather than replacing conventional floorplanning methods.

\subsection{Experimental Setup}
\label{subsec:experimental_setup}

We conducted experiments using two approaches. First, we utilize Unsloth \cite{unsloth}, an efficient framework for locally fine-tuning open-source LLMs. Second, we leverage the OpenAI API to fine-tune a proprietary LLM, specifically GPT4o-mini \cite{gpt4omini}.

\subsubsection{Dataset Specifications}
\label{subsubsec:dataset_spec}

\begin{itemize}
  \item \textbf{Module Counts:} Experiments were performed on datasets with 16 and 24 modules. These counts were inspired by the ami33 and ami49 circuits from the MCNC benchmark, originally containing 33 and 49 modules, respectively. To align with our model requirements, we combined two to four modules with identical widths or heights into single modules.
  \item \textbf{Dataset Sizes:} The training datasets comprised 80,000 and 120,000 optimal floorplans for the 16-module and 24-module cases, respectively. Larger module counts necessitated larger datasets to enhance fine-tuning performance.
  \item \textbf{Test Samples:} Each model was evaluated on 50 unseen test samples generated using the same recursive slicing and encoding methodology as the training data.
\end{itemize}

\subsubsection{Metrics for Evaluation}
\label{subsubsec:evaluation_metrics}

\begin{itemize}
  \item \textbf{Success Rate (\(S\)):} Percentage of test samples yielding a legal slicing tree.
  \item \textbf{Optimal Result Rate (\(O\)):} Percentage of floorplans with zero dead space, indicating an optimal solution.
  \item \textbf{Dead Space Ratio (\(D\)):} Ratio of dead space to the summation of all module areas in the floorplan.
\end{itemize}

The metrics are mathematically defined as:

\begin{equation}
  S = \frac{N_{\text{legal}}}{N_{\text{total}}} \times 100\%
  \label{eq:success_rate}
\end{equation}

\begin{equation}
  O = \frac{N_{\text{optimal}}}{N_{\text{total}}} \times 100\%
  \label{eq:optimal_rate}
\end{equation}

\begin{equation}
  D = \frac{\sum_{i=1}^{n-1} \text{dead\_space}(p_i, p_{i+1})}{\sum_{i=1}^{n} \left( w_i \times h_i \right)}
  \label{eq:dead_space_ratio}
\end{equation}

where \(N_{\text{legal}}\) is the number of test samples with legal slicing trees, \(N_{\text{total}}\) is the total number of test samples, \(N_{\text{optimal}}\) is the number of optimal floorplans, and \(\text{dead\_space}(p_i, p_{i+1})\) is the dead space between two adjacent modules and can be calculated using Equations~\ref{eq:dead_space}.

\subsubsection{Model Fine-tuning Details}
\label{subsubsec:model_finetuning_details}

\begin{itemize}
  \item \textbf{Local LLM Fine-tuning:} Models fine-tuned locally include LLaMA 3.1 (8B) \cite{meta_llama_3_1_2024}, LLaMA 3.2 (3B) \cite{meta_llama_3_2_connect_2024}, Mistral v0.3 (7B) \cite{jiang2023mistral}, and Phi-4 (13B) \cite{abdin2024phi}. The Unsloth framework was chosen for its efficiency, achieving 2-5x faster fine-tuning with 70\% less memory usage compared to traditional frameworks. Fine-tuning was performed on a single NVIDIA GeForce RTX 4070 Ti Super with 16 GB of VRAM over 200 epochs, completing in approximately 30 minutes.

  \item \textbf{OpenAI API Fine-tuning:} Due to hardware constraints that made it challenging to fine-tune larger LLMs locally, we opted to fine-tune OpenAI's GPT4o-mini. Initial fine-tuning on the 16-module dataset yielded promising results. Although we expanded the dataset size to test the capability of LLMs locally, budget limitations restricted us to re-fine-tuning only the 16-module model, resulting in 6,253 samples for GPT4o-mini.
\end{itemize}

\section{Experimental Results}
\label{sec:experimental_results}
In this section, we present our experimental results for two approaches: (1) using Unsloth \cite{unsloth}, an efficient framework for fine-tuning open-source LLMs locally, and (2) leveraging the OpenAI API to fine-tune a proprietary LLM for the floorplanning task. The experiments are conducted on datasets containing 16 and 24 modules. The choice of these module counts was inspired by the ami33 and ami49 circuits from the MCNC benchmark, which originally contain 33 and 49 modules, respectively. To align with the requirements of our fine-tuned models, we combine two to four modules (with the same width or height) into a single module. However, since benchmark data does not always form rectangles, our experiments are based on custom benchmarks that will be made publicly available in the future.

\subsection{Quantitative Results}
In this section, we present quantitative results for the success rate, optimal rate, and average dead space. To validate the hypothesis that LLMs can efficiently solve floorplanning problems for specific module counts based on the training data, we evaluate the performance of the fine-tuned models on 50 unseen test samples. For models fine-tuned on the 16- and 24-module datasets, we evaluate their performance by testing within a ±3 module range to assess the models' generalization ability to varying module counts. For each test sample, the LLM generates five outputs, and the best result is selected for evaluation. The equations used to compute these metrics are provided in Equations~\ref{eq:success_rate} to~\ref{eq:dead_space_ratio}.

\subsubsection{Success Rate}
\begin{figure}[h]
  \centering
  \begin{subfigure}[b]{0.45\textwidth}
    \centering
    \includegraphics[width=\textwidth]{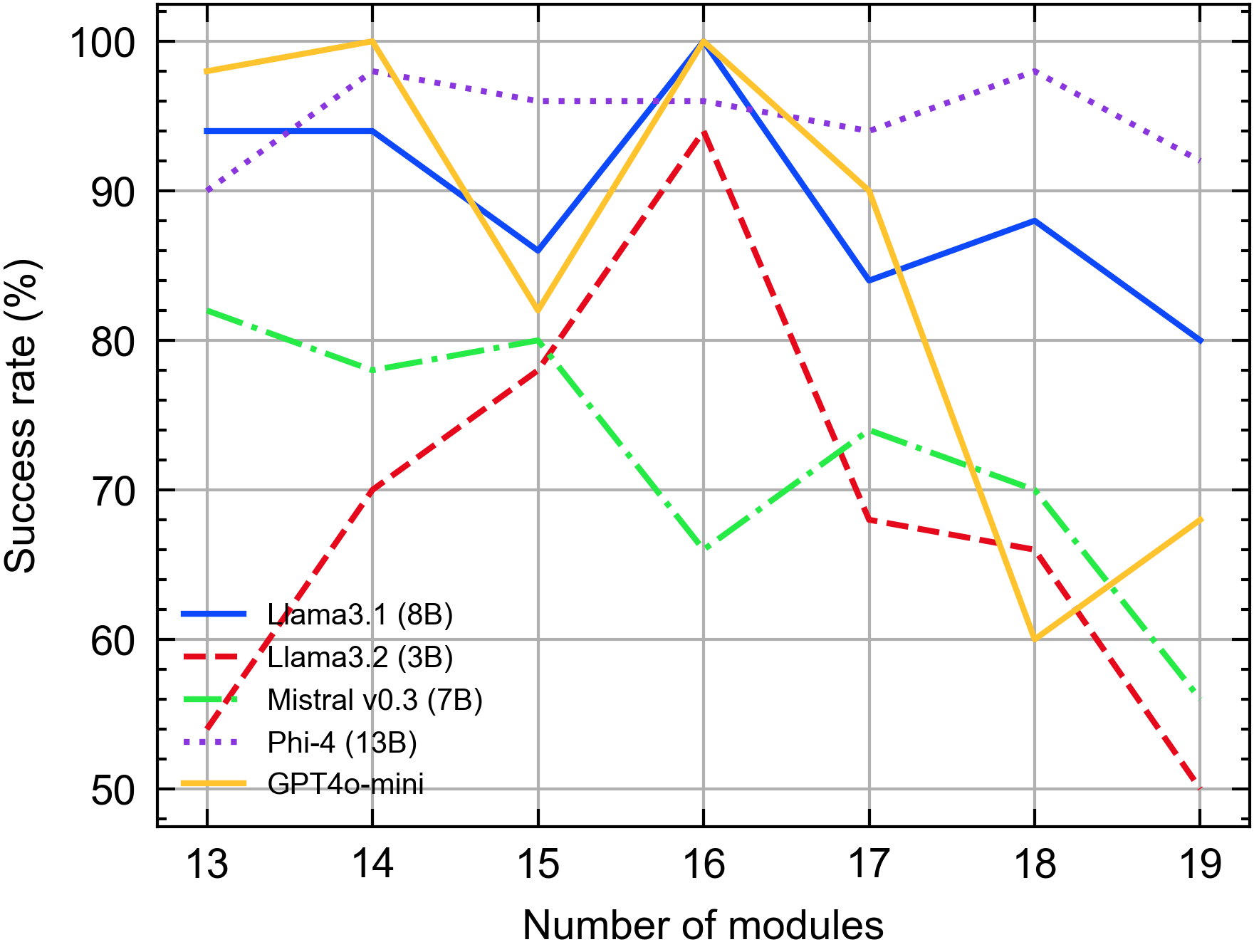}
    \caption{16-module scenario}
    \Description{The success rate of 16-module fine-tuned LLMs for different module counts. The results show that GPT4o-mini, Phi-4 (13B), and Llama3.1 (8B) achieve high success rates, while Mistral v0.3 (7B) and Llama3.2 (3B) perform poorly. For Mistral v0.3 (7B), the success rate decreases at 16 modules strangely.}
    \label{fig:success_rate_1}
  \end{subfigure}
  \hfill
  \begin{subfigure}[b]{0.45\textwidth}
    \centering
    \includegraphics[width=\textwidth]{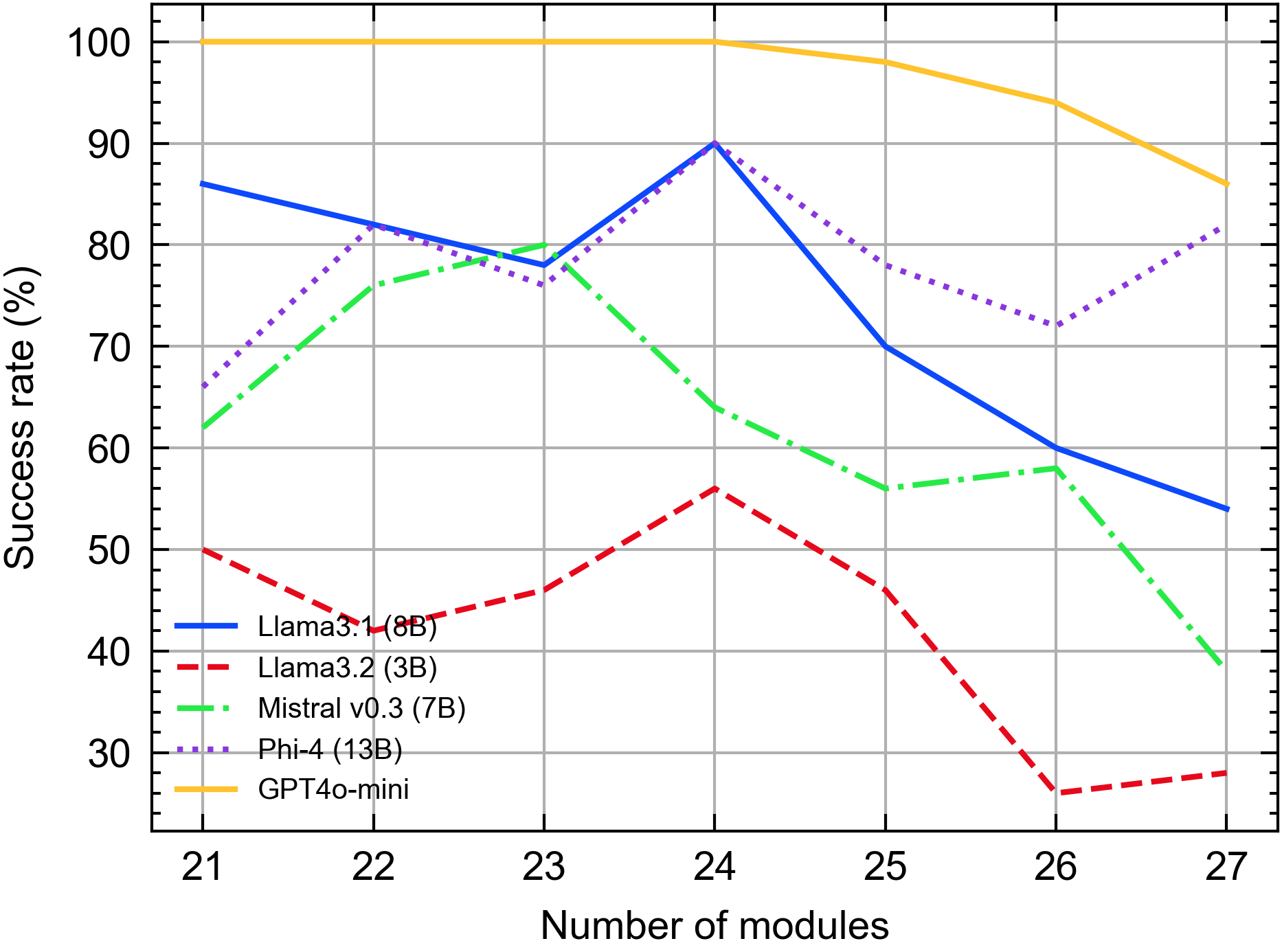}
    \caption{24-module scenario}
    \Description{The success rate of 24-module fine-tuned LLMs for different module counts. The results show that GPT4o-mini, Phi-4 (13B), and Llama3.1 (8B) achieve high success rates, while Mistral v0.3 (7B) and Llama3.2 (3B) perform poorly.}
    \label{fig:success_rate_2}
  \end{subfigure}
  \caption{Success rate of fine-tuned LLMs for different module counts.}
  \Description{Success rate for fine-tuned LLMs for different module counts.}
  \label{fig:success_rate}
\end{figure}

Figure~\ref{fig:success_rate} illustrates the success rate for the different fine-tuned LLMs, which is defined as the percentage of test samples that produce a legal slicing tree. In the 16-module scenario (Figure~\ref{fig:success_rate_1}), GPT4o-mini, Phi-4 (13B), and Llama3.1 (8B) achieve high success rates—with nearly 100\% at 16 modules—while Mistral v0.3 (7B) shows a noticeable decline at 16 modules, failing to meet our hypothesis that the model should complete the specified module count. Similarly, in the 24-module scenario (Figure~\ref{fig:success_rate_2}), GPT4o-mini demonstrates near-perfect performance at lower module counts with slight declines as the module count increases. In contrast, Phi-4 (13B) and Llama3.1 (8B) perform well for the 16-module case, whereas Mistral v0.3 (7B) and Llama3.2 (3B) achieve only around a 60\% success rate. Overall, these results indicate that GPT4o-mini, Phi-4 (13B), and Llama3.1 (8B) are the most robust models for generating legal slicing trees in our floorplanning tasks.

\subsubsection{Optimal Rate}
\begin{figure}[h]
  \centering
  \begin{subfigure}[b]{0.45\textwidth}
    \centering
    \includegraphics[width=\textwidth]{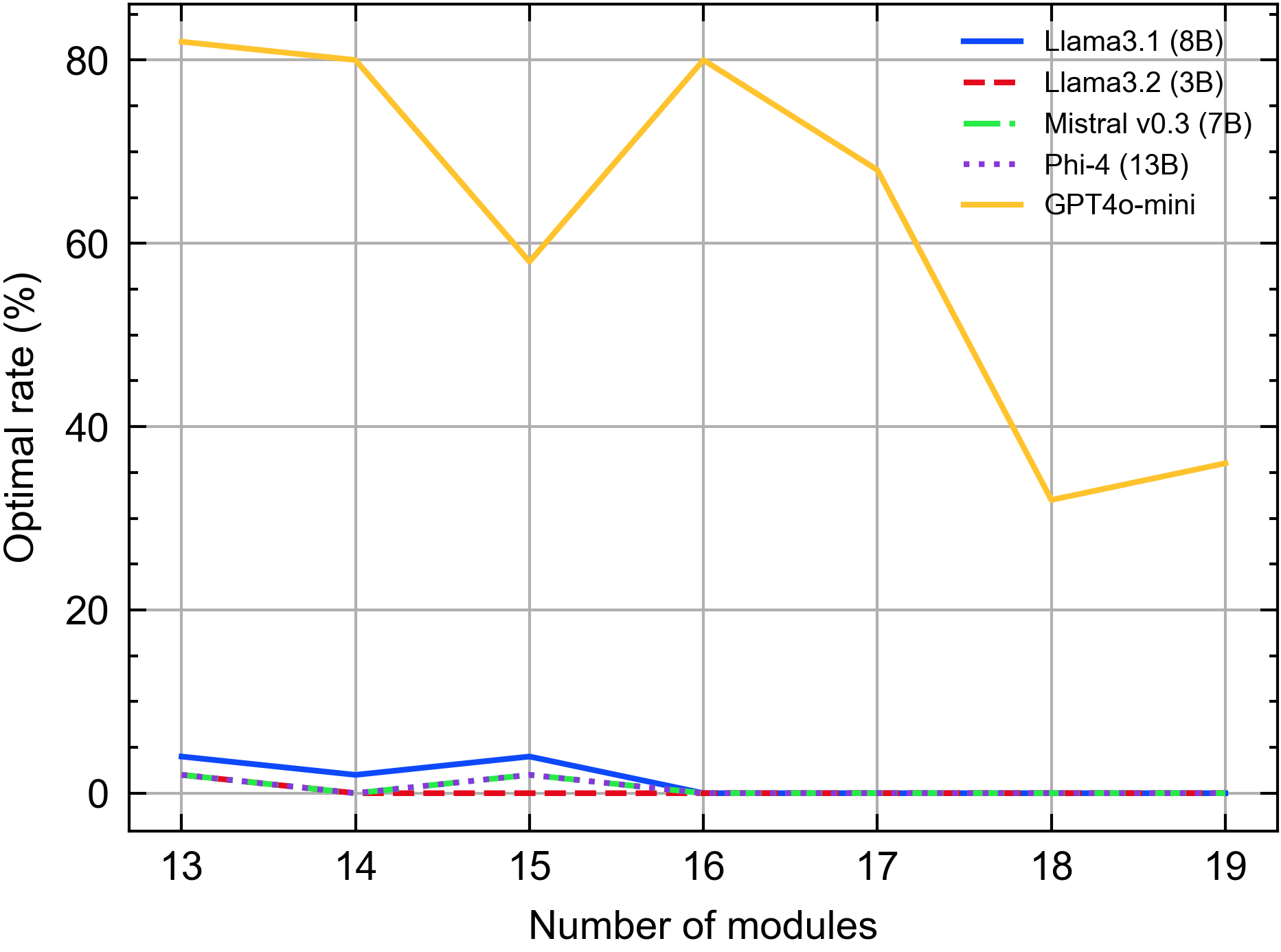}
    \caption{16-module scenario}
    \Description{The optimal rate of 16-module fine-tuned LLMs for different module counts. The results show that GPT4o-mini achieves significantly higher optimal rates compared to other models. At low module counts, Llama3.1 (8B) achieves optimal rates up to 4\%, while Llama3.2 (3B), Mistral v0.3 (7B), and Phi-4 (13B) achieve negligible rates.}
    \label{fig:opt_rate_1}
  \end{subfigure}
  \hfill
  \begin{subfigure}[b]{0.45\textwidth}
    \centering
    \includegraphics[width=\textwidth]{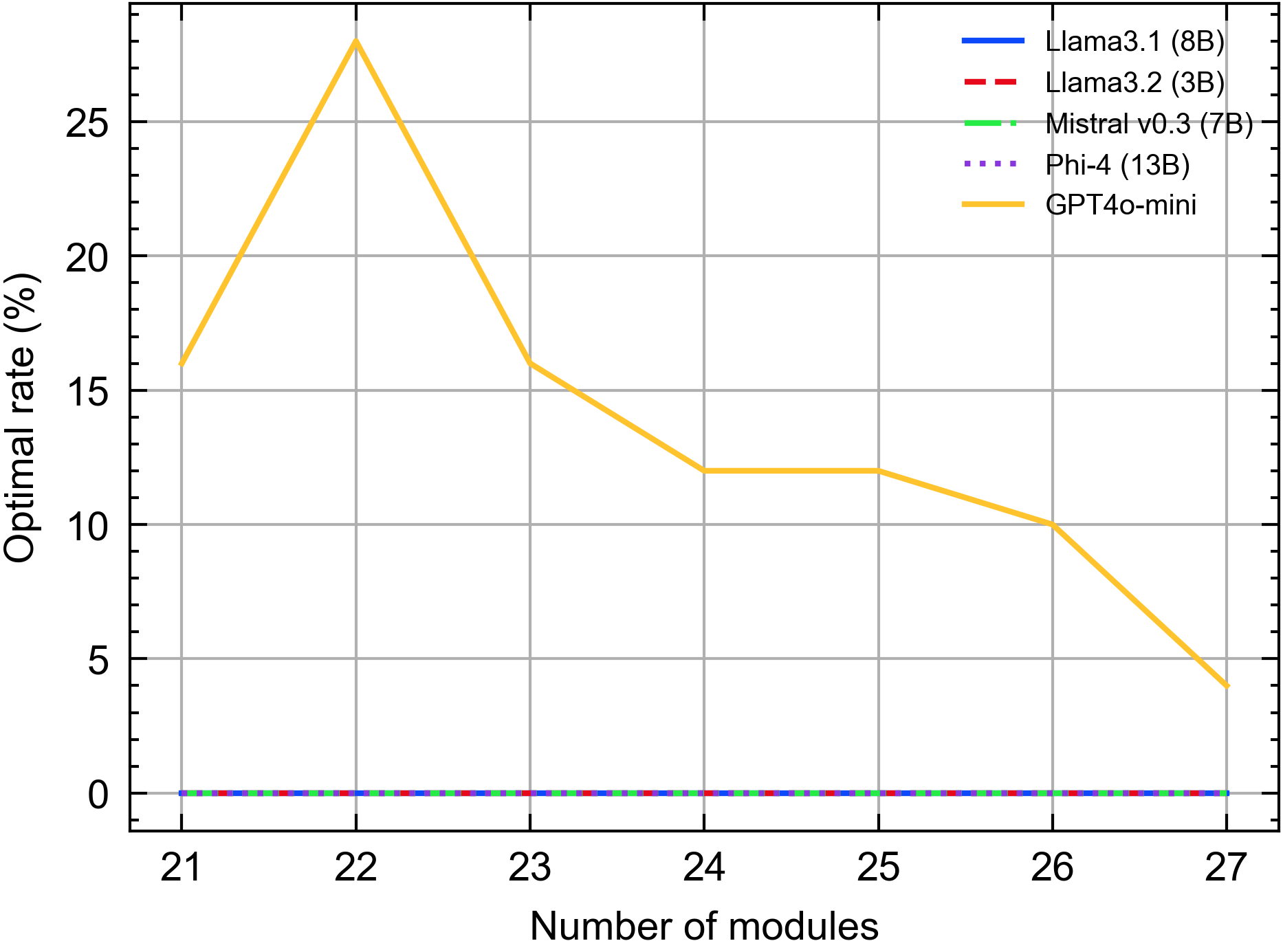}
    \caption{24-module scenario}
    \Description{The optimal rate of 24-module fine-tuned LLMs for different module counts. The results show that GPT4o-mini achieves optimal rates up to 28\% at 22 modules, while the other models consistently produce 0\% optimal results.}
    \label{fig:opt_rate_2}
  \end{subfigure}
  \caption{Optimal rate of fine-tuned LLMs for different module counts.}
  \Description{Optimal rate for fine-tuned LLMs for different module counts.}
  \label{fig:opt_rate}
\end{figure}

Figure~\ref{fig:opt_rate} shows the optimal rate for two fine-tuned LLMs, defined as the percentage of floorplanning results that achieve an optimal floorplan. In the 16-module scenario (Figure~\ref{fig:opt_rate_1}), GPT4o-mini exhibits substantially higher optimal rates (ranging from 57\% to 82\%) compared to the other models. In contrast, Llama3.1 (8B) records optimal rates up to only 4\% for some module counts, while Llama3.2 (3B), Mistral v0.3 (7B), and Phi-4 (13B) achieve negligible rates (mostly between 0\% and 2\%). In the 24-module scenario (Figure~\ref{fig:opt_rate_2}), GPT4o-mini attains optimal rates up to 28\% at 22 modules, though it decreases to 12\% at 24 modules. The other models consistently produce 0\% optimal results. These findings indicate that GPT4o-mini is significantly more effective at generating optimal floorplanning solutions, especially as the number of modules increases, though its performance decreases with higher module counts, reflecting inherent limitations.

\subsubsection{Average Dead Space}
\begin{figure}[h]
  \centering
  \begin{subfigure}[b]{0.45\textwidth}
    \centering
    \includegraphics[width=\textwidth]{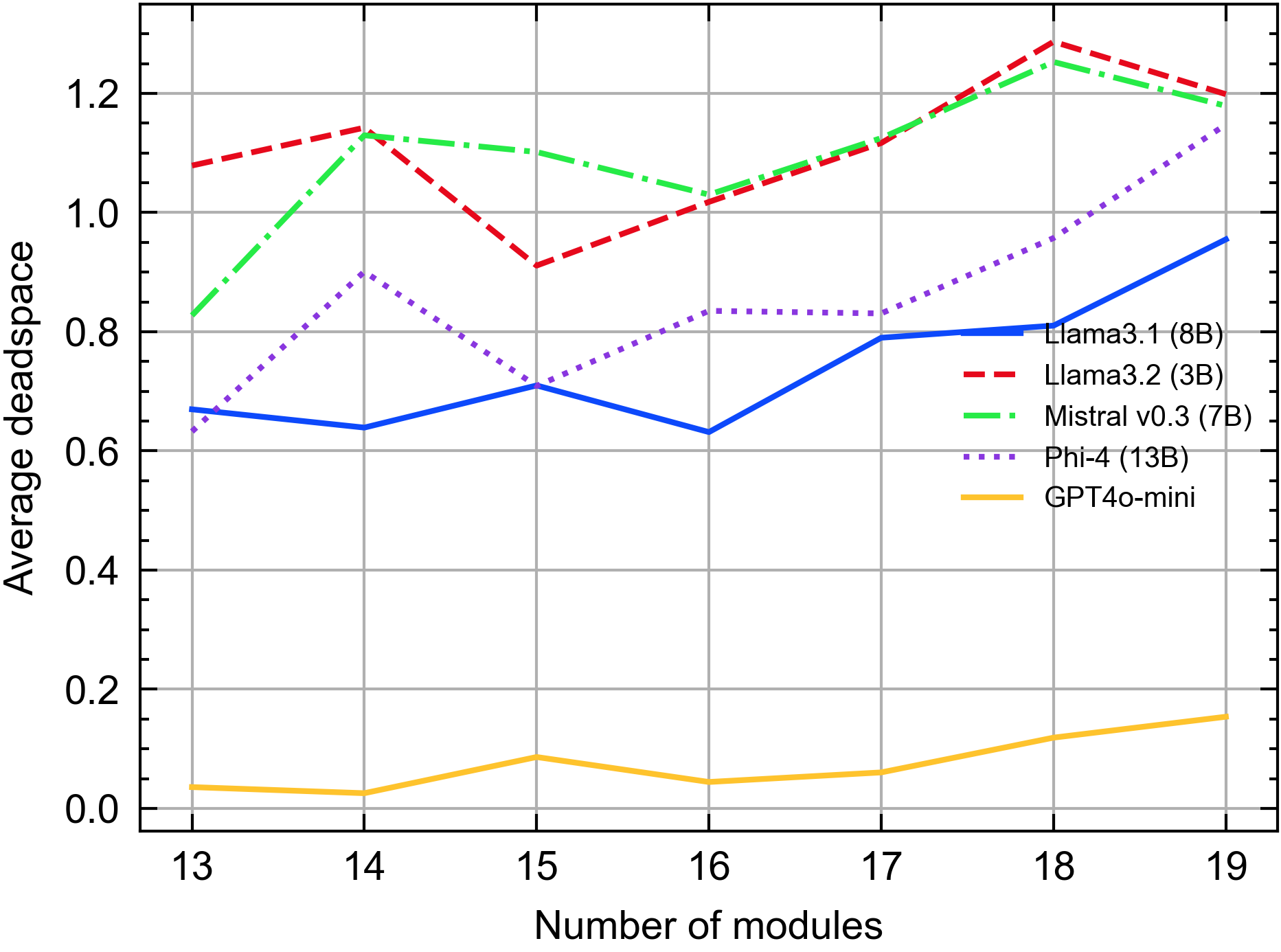}
    \caption{16-module scenario}
    \Description{The average dead space of fine-tuned LLMs for different module counts. The results show that GPT4o-mini achieves very low average dead space values, while the other models exhibit significantly higher values.}
    \label{fig:avg_dead_1}
  \end{subfigure}
  \hfill
  \begin{subfigure}[b]{0.45\textwidth}
    \centering
    \includegraphics[width=\textwidth]{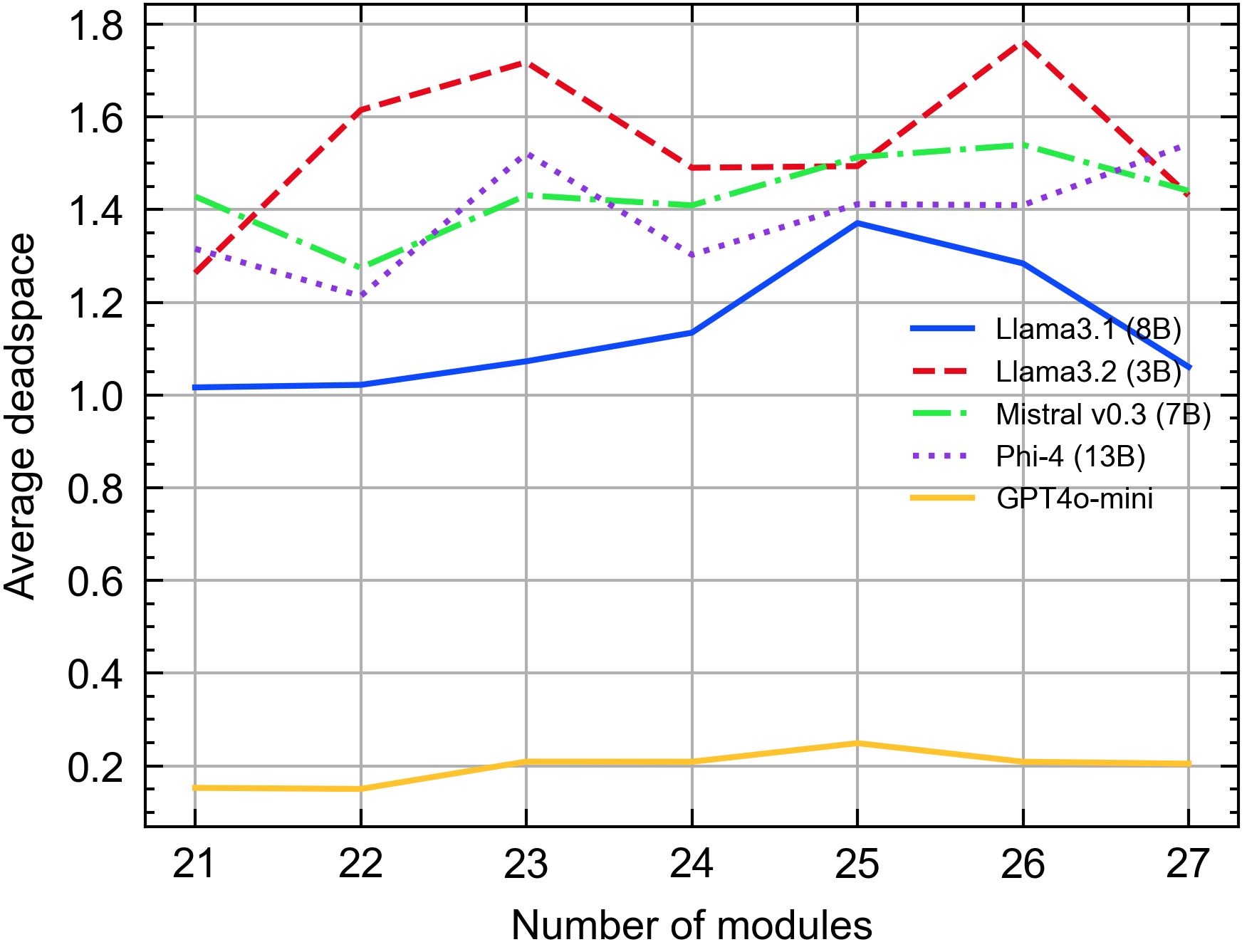}
    \caption{24-module scenario}
    \Description{The average dead space of fine-tuned LLMs for different module counts. The results show that GPT4o-mini maintains low average dead space values, while the other models exhibit significantly higher values.}
    \label{fig:avg_dead_2}
  \end{subfigure}
  \caption{Average dead space of fine-tuned LLMs for different module counts.}
  \Description{Average dead space for fine-tuned LLMs for different module counts.}
  \label{fig:avg_dead}
\end{figure}

Figure~\ref{fig:avg_dead} presents the average dead space values for the legal slicing trees generated by the fine-tuned LLMs. (Only legal slicing trees are considered in the computation.) Lower values indicate a more efficient floorplan with less unused space. In the 16-module scenario (Figure~\ref{fig:avg_dead_1}), GPT4o-mini achieves very low average dead space values (ranging from 0.03 to 0.15), whereas the other models exhibit significantly higher values: Llama3.1 (8B) ranges from 0.63 to 0.95, Llama3.2 (3B) from 0.91 to 1.20, Mistral v0.3 (7B) from 0.83 to 1.18, and Phi-4 (13B) from 0.63 to 1.15. In the 24-module scenario (Figure~\ref{fig:avg_dead_2}), GPT4o-mini maintains low average dead space values (between 0.15 and 0.25), while Llama3.1 (8B) ranges from 1.02 to 1.37, Llama3.2 (3B) from 1.26 to 1.76, Mistral v0.3 (7B) from 1.27 to 1.54, and Phi-4 (13B) from 1.21 to 1.54. These results demonstrate that GPT4o-mini not only generates legal slicing trees at a higher success rate but also produces significantly lower dead space, indicating a more efficient allocation of chip area.

\subsection{Discussion}
Overall, our experimental results demonstrate that GPT4o-mini outperforms the other fine-tuned LLMs by a significant margin. In the 16-module scenario (module counts 13-19), GPT4o-mini achieves high optimal rates (ranging from 57\% to 82\%) and exceptionally low average dead space values (0.03 to 0.15), thereby generating legal slicing trees with minimal unused space. In contrast, while some models (e.g., Llama3.1 (8B)) achieve high success rates, their optimal rates remain very low (mostly between 0\% and 4\%), and they yield significantly higher average dead space values, indicating less efficient floorplanning.

In the 24-module scenario (module counts 21-27), only GPT4o-mini is capable of generating optimal slicing trees—albeit with reduced success rates (4\% to 28\%)—while the other models fail to produce any optimal outputs. These findings clearly indicate that GPT4o-mini is the most robust and efficient model for generating legal slicing trees in our floorplanning tasks. Although more powerful models, such as GPT4o \cite{gpt4o}, might further enhance optimal performance, the high cost of fine-tuning and our limited budget prevented further exploration in that direction. Nonetheless, our results validate the hypothesis that LLMs can perform floorplanning in a manner analogous to human subitizing.

While our current experiments fine-tuned LLMs only for 16- and 24-module datasets, we believe that performance can be further improved by providing additional data, offering more detailed instructions, or employing more powerful models. It is important to note that our evaluation focused solely on floorplanning with optimal results. In practice, the floorplanning problem is more complex, involving additional constraints such as wire length minimization, thermal management, and power consumption. Future research should address these factors to develop more comprehensive and practical solutions for VLSI floorplanning. Moreover, we posit that fine-tuned LLMs can be applied to other VLSI design challenges, such as partitioning or timing analysis, serving as a complementary tool to guide traditional algorithms toward better solutions.

\section{Conclusion}
\label{sec:conclusion}

We introduced an innovative approach to solving the floorplanning problem in VLSI design by leveraging fine-tuned Large Language Models (LLMs). We developed an efficient representation and a novel method for generating high-quality datasets necessary for effective LLM fine-tuning. Our experimental evaluation revealed that fine-tuned LLMs, particularly GPT4o-mini, achieve high success and optimal rates while producing significantly lower average dead space, resulting in more efficient floorplans compared to traditional and other ML-based methods. These findings support our hypothesis that LLMs can perform floorplanning tasks in a manner akin to human subitizing, demonstrating their potential as powerful tools for addressing complex optimization problems in VLSI design.

Despite these promising results, the performance of GPT4o-mini diminishes at 24-module scenario, highlighting the need for further research to enhance scalability and incorporate additional design constraints. Future work should explore the integration of advanced LLM architectures and the inclusion of comprehensive optimization objectives—such as wire length minimization, thermal management, and power consumption—to develop more robust and practical floorplanning solutions. Our findings pave the way for more intelligent and efficient design automation processes in VLSI design.

\begin{acks}
  We would like to express our gratitude to OpenAI for providing access to ChatGPT, which significantly contributed to enhancing the clarity, coherence, and overall quality of this paper. The assistance in refining technical descriptions, structuring content, and generating insights was invaluable throughout the research and writing process.
\end{acks}

\bibliographystyle{ACM-Reference-Format}
\bibliography{Subitizing-Inspired_Large_Language_Models_for_Floorplanning}

\appendix

\section{Floorplanning Dataset}
\begin{table}[H]
  \centering
  \caption{Floorplanning Dataset Example}
  \begin{tabular}{p{0.95\linewidth}}
    \hline
    \textbf{Instruction}                                                                                                                \\
    \textbf{1. Input:}                                                                                                                  \\
    A string containing \textbf{n rectangular modules} defined as \texttt{P\_i(width, height)}, separated by semicolons.                \\
    \textbf{Example:}                                                                                                                   \\
    P\_5(5412,522);P\_83(3442,1961);P\_87(1970,1961)                                                                                    \\
    \\
    \textbf{2. Output:}                                                                                                                 \\
    A slicing tree in \textbf{post-order traversal} format where:                                                                       \\
    \quad \textbf{Leaf nodes:} Module names (\texttt{P\_i}).                                                                            \\
    \quad \textbf{Internal nodes:} Horizontal (\texttt{H}) for vertical stacking, and Vertical (\texttt{V}) for side-by-side placement. \\
    The solution must minimize the \textbf{total layout area}.                                                                          \\
    \textbf{Example Output:}                                                                                                            \\
    P\_83;P\_87;V;P\_5;H                                                                                                                \\
    \\
    \textbf{3. Step-by-Step Process:}                                                                                                   \\
    \textbf{Step 1: Parse the Input}                                                                                                    \\
    - Extract module dimensions and labels from the input string.                                                                       \\
    - Store them in a list of tuples.                                                                                                   \\
    \\
    \textbf{Step 2: Minimize the Total Area}                                                                                            \\
    - Explore different configurations for placing the modules using \texttt{H} and \texttt{V} cuts.                                    \\
    - Compute the layout dimensions for each configuration and choose the one that minimizes the total area (\(width \times height\)).  \\
    \\
    \textbf{Step 3: Construct the Slicing Tree}                                                                                         \\
    - Build a binary slicing tree where:                                                                                                \\
    \quad - \textbf{Leaf nodes} represent the modules.                                                                                  \\
    \quad - \textbf{Internal nodes (\(H\) or \(V\))} represent the selected cuts.                                                       \\
    \quad - The left (bottom) module will be the left node.                                                                             \\
    \quad - The right (top) module will be the right node.                                                                              \\
    \\
    \textbf{Step 4: Return Post-Order Traversal Result}                                                                                 \\
    - Traverse the slicing tree in post-order (left subtree \(\rightarrow\) right subtree \(\rightarrow\) root) and return the result.  \\
    \hline
    \textbf{Prompt}                                                                                                                     \\
    P\_18(4455,4857);P\_22(1984,8286);P\_24(3376,8286);P\_28(619,4177);P\_41(402,4177);P\_47(6078,2990);                                \\
    P\_51(315,4177);P\_53(1057,2238);P\_62(4455,3429);P\_66(10866,2551);P\_67(10866,39);P\_73(1057,1939);                               \\
    P\_77(1832,4177);P\_78(6078,1187);P\_86(1051,8286);P\_87(563,4177)                                                                  \\
    \hline
  \end{tabular}
  \label{tab:floorplanning_instructions}
\end{table}

Table~\ref{tab:floorplanning_instructions} presents an example of the floorplanning dataset, which includes both the instruction and the prompt. The instruction specifies the input and output format and details the step-by-step process for generating the slicing tree, while the prompt provides a set of modules with their respective width and height. The model uses this information to generate an optimal floorplan. We generated 80,000 and 120,000 samples for the 16-module and 24-module datasets, respectively, using this format.

\section{Fine-tuning Configuration and Loss Values}
Here, we provide detailed configuration settings and loss values for the local LLM fine-tuning and OpenAI API fine-tuning experiments. For local LLM, we used the Unsloth framework with the LLaMA 3.1 (8B), LLaMA 3.2 (3B), Mistral v0.3 (7B), and Phi-4 (13B) models. For OpenAI API fine-tuning, we utilized the GPT4o-mini model. Note that for OpenAI API fine-tuning, 16-module fine-tuning is a little bit different to other models since we did not have enough budget to re-fine-tune the model. The configuration settings and loss values are presented below.

\subsection{Local LLM Fine-tuning}
\begin{figure}[h]
  \centering
  \begin{subfigure}[b]{0.45\textwidth}
    \centering
    \includegraphics[width=\textwidth]{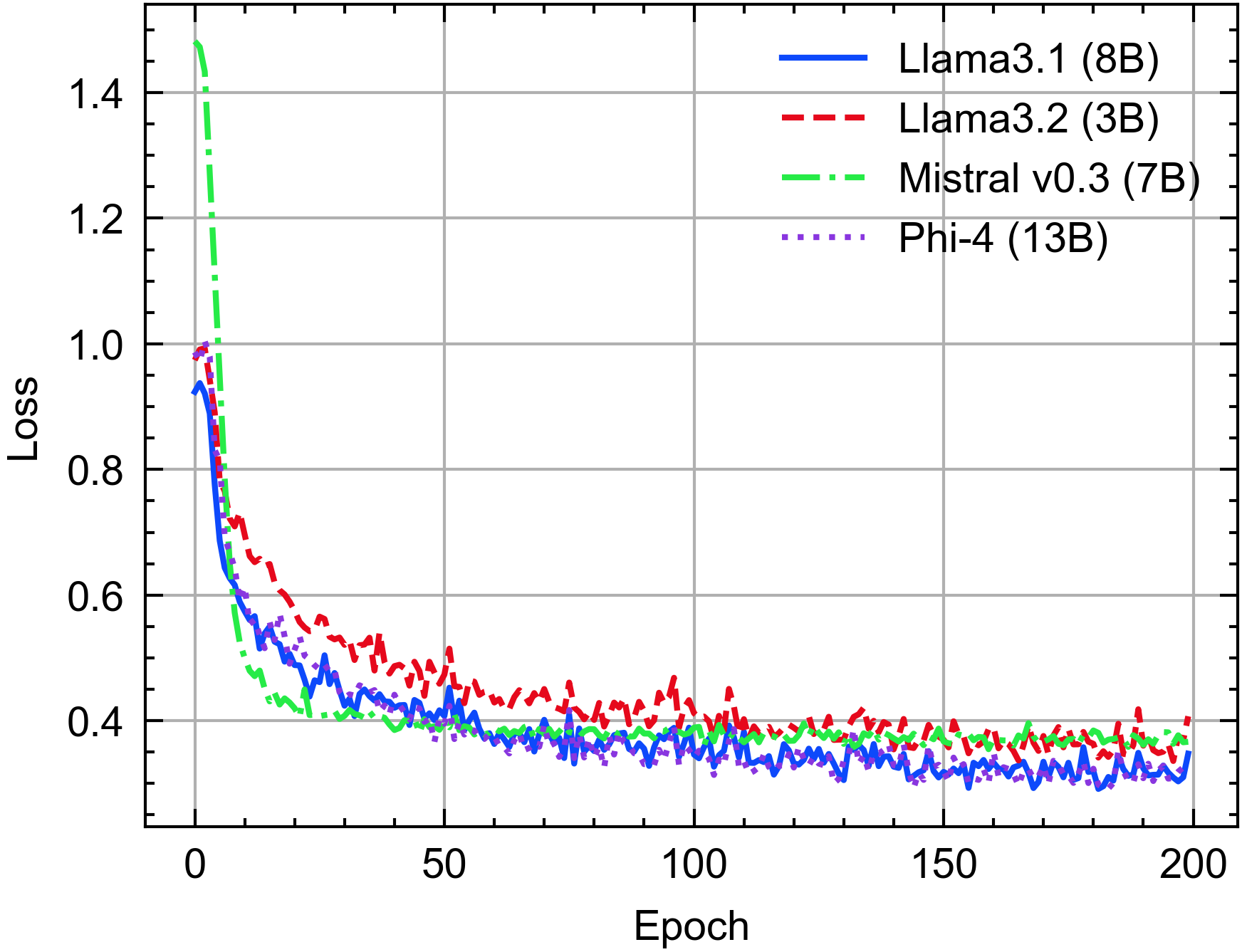}
    \caption{16-module scenario}
    \Description{The local LLM fine-tuning loss values for the 16-module scenario. The loss values indicate the convergence of the fine-tuning process, with lower values suggesting better model performance.}
    \label{fig:local_loss_16}
  \end{subfigure}
  \hfill
  \begin{subfigure}[b]{0.45\textwidth}
    \centering
    \includegraphics[width=\textwidth]{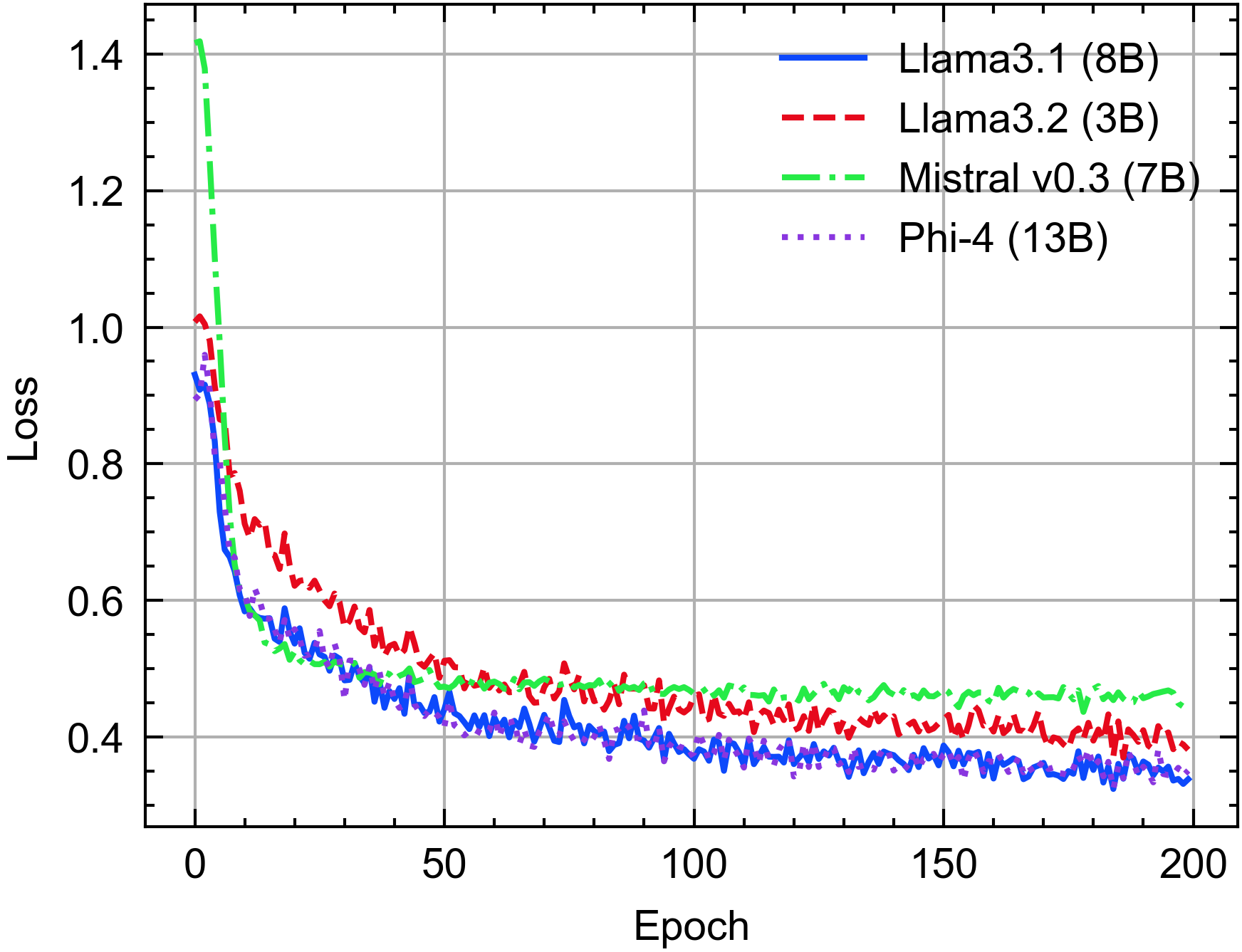}
    \caption{24-module scenario}
    \Description{The local LLM fine-tuning loss values for the 24-module scenario. The loss values indicate the convergence of the fine-tuning process, with lower values suggesting better model performance.}
    \label{fig:local_loss_24}
  \end{subfigure}
  \caption{Local LLM fine-tuning loss values for 16 and 24 modules.}
  \Description{Local LLM fine-tuning loss values for 16 and 24 modules.}
  \label{fig:local_loss}
\end{figure}

Table~\ref{tab:local_experiment_config} shows the configuration settings for the local LLM fine-tuning experiments. There are some other parameter can set in Unsloth, but we only show the most important ones, others remain default. The loss values for the 16-module and 24-module scenarios are depicted in Figure~\ref{fig:local_loss}. The loss values indicate the convergence of the fine-tuning process, with lower values suggesting better model performance.

\begin{table}[H]
  \centering
  \begin{minipage}{0.45\linewidth}
    \begin{table}[H]
      \centering
      \caption{Experiment Configuration Settings}
      \begin{tabular}{ll}
        \hline
        \textbf{Configuration Parameter} & \textbf{Value} \\ \hline
        max sequence length              & 2048           \\
        load in 4-bit                    & True           \\
        warmup steps                     & 5              \\
        learning rate                    & 2e-4           \\
        batch size                       & 2              \\
        optimizer                        & AdamW 8-bit    \\
        epochs                           & 200            \\
        learning rate scheduler          & linear         \\
        seed                             & 3407           \\ \hline
      \end{tabular}
      \label{tab:local_experiment_config}
    \end{table}
  \end{minipage}
\end{table}

\subsection{OpenAI API Fine-tuning}
\begin{figure}[h]
  \centering
  \begin{subfigure}[b]{0.45\textwidth}
    \centering
    \includegraphics[width=\textwidth]{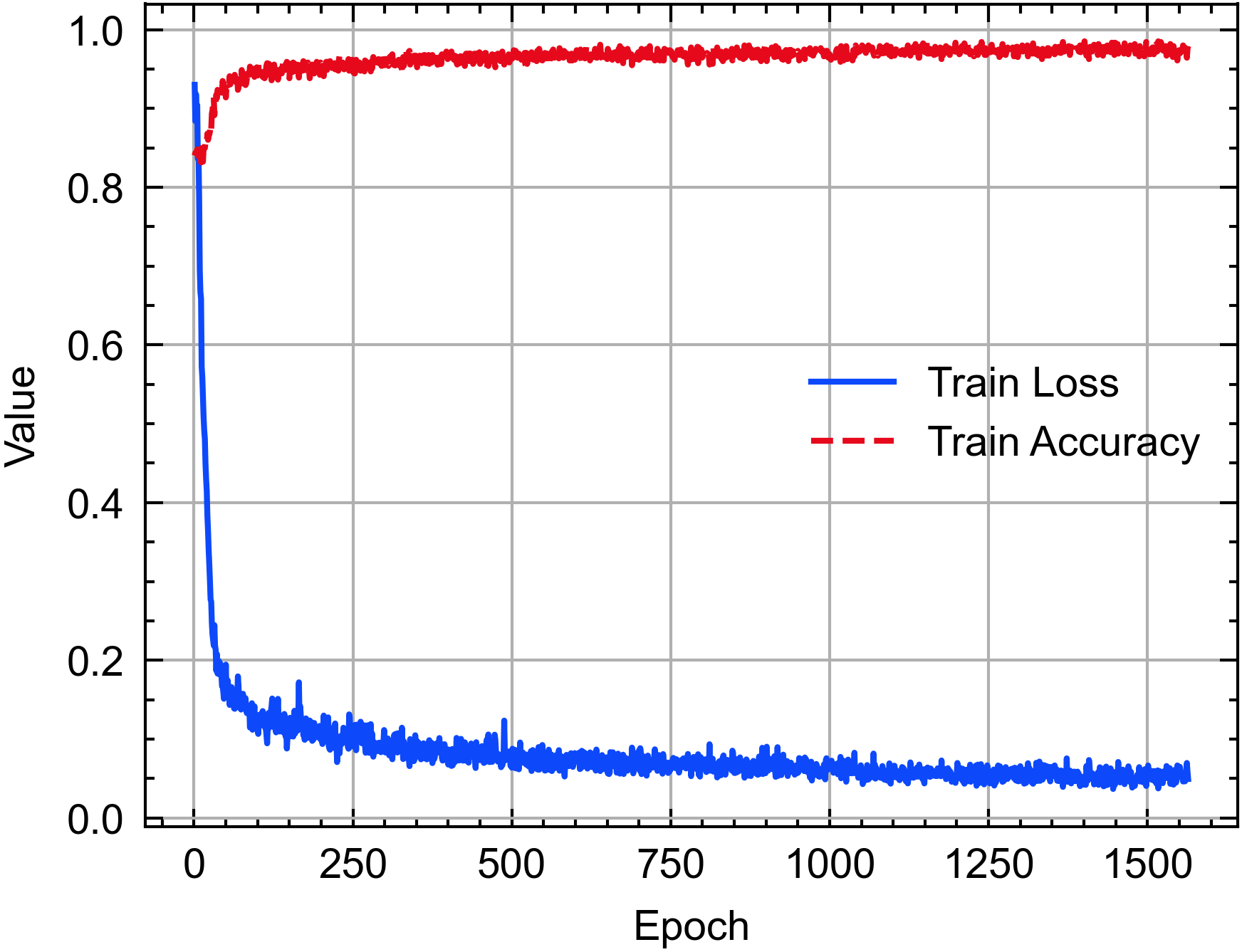}
    \caption{16-module scenario}
    \Description{OpenAI API fine-tuning loss curves for the 16-module dataset.}
    \label{fig:openai_loss_16}
  \end{subfigure}
  \hfill
  \begin{subfigure}[b]{0.45\textwidth}
    \centering
    \includegraphics[width=\textwidth]{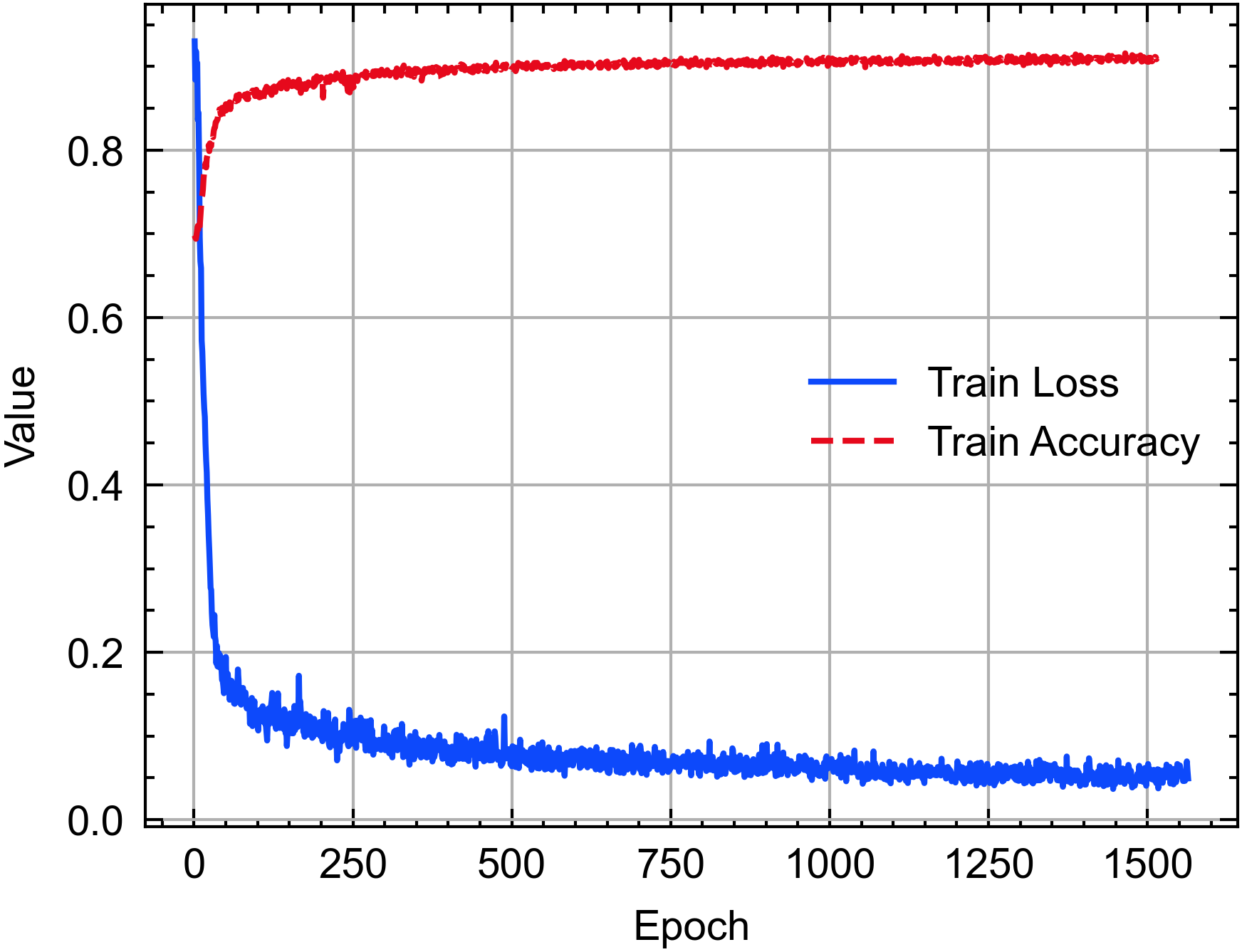}
    \caption{24-module scenario}
    \Description{OpenAI API fine-tuning loss curves for the 24-module dataset.}
    \label{fig:openai_loss_24}
  \end{subfigure}
  \caption{Loss curves during OpenAI API fine-tuning for the 16-module and 24-module datasets.}
  \Description{OpenAI API fine-tuning loss curves for 16-module and 24-module datasets}
  \label{fig:openai_loss}
\end{figure}

Table~\ref{tab:api_experiment_config_16} and Table~\ref{tab:api_experiment_config_24} summarize the configuration settings for the OpenAI API fine-tuning experiments on the 16-module and 24-module datasets, respectively. The datasets for 16- and 24-module fine-tuning were 20MB and 200MB, consuming approximately 17.3 and 73.9 million tokens, respectively, incurring a total cost of \$220. The fine-tuned model demonstrated superior performance compared to local LLMs, particularly in generating valid and optimized floorplanning results. Despite this limitation, GPT4o-mini demonstrated superior performance compared to locally fine-tuned models. The total training time for the 16-module dataset was approximately 45 minutes, whereas fine-tuning the 24-module dataset required around 3 hours due to its larger size. For the 16-module experiments, the model was fine-tuned for 3 epochs over 1560 steps; in contrast, the 24-module dataset was fine-tuned for 1 epoch over 1561 steps. Figure~\ref{fig:openai_loss} displays the loss curves for both scenarios, demonstrating the convergence of the fine-tuning process—lower loss values indicate better model performance.

\begin{table}[H]
  \centering
  \begin{minipage}{0.45\linewidth}
    \centering
    \caption{16-Module Fine-tuning Configuration Settings}
    \begin{tabular}{ll}
      \hline
      \textbf{Parameter} & \textbf{Value}         \\ \hline
      Training Method    & Supervised             \\
      Base Model         & gpt-4o-mini-2024-07-18 \\
      Epochs             & 3                      \\
      Batch Size         & 12                     \\
      LR Multiplier      & 1.8                    \\
      Seed               & 1760895552             \\ \hline
    \end{tabular}
    \label{tab:api_experiment_config_16}
  \end{minipage}%
  \hfill
  \begin{minipage}{0.45\linewidth}
    \centering
    \caption{24-Module Fine-tuning Configuration Settings}
    \begin{tabular}{ll}
      \hline
      \textbf{Parameter} & \textbf{Value}         \\ \hline
      Training Method    & Supervised             \\
      Base Model         & gpt-4o-mini-2024-07-18 \\
      Epochs             & 1                      \\
      Batch Size         & 66                     \\
      LR Multiplier      & 1.8                    \\
      Seed               & 2118753743             \\ \hline
    \end{tabular}
    \label{tab:api_experiment_config_24}
  \end{minipage}
\end{table}

\clearpage

\section{Original Data for Dead Space Ratios}
\begin{figure}[h]
  \centering
  \includegraphics[width=\textwidth]{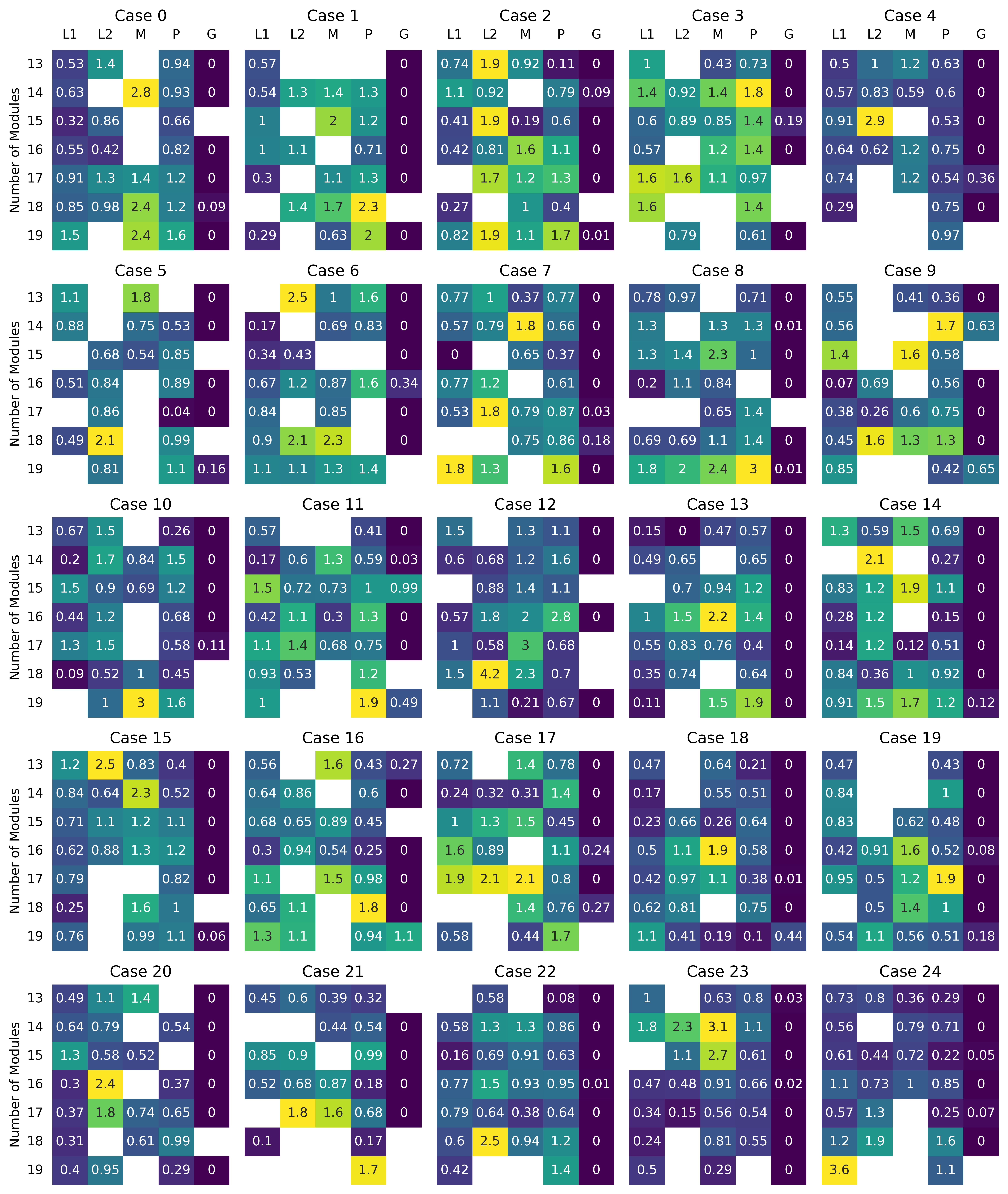}
  \caption{Original data for dead space ratios for the 16-module model (Cases 0 to 24). Brighter colors indicate lower dead space ratios. The labels "L1", "L2", "M", "P", and "G" represent "Llama3.1 (8B)", "Llama3.2 (3B)", "Mistral v0.3 (7B)", "Phi-4 (13B)", and "GPT4o-mini", respectively.}
  \Description{Plot showing dead space ratios for the 16-module dataset from cases 0 to 24 across different models, where brighter colors indicate lower dead space ratios.}
  \label{fig:app_ori_16_0}
\end{figure}

\begin{figure}[h]
  \centering
  \includegraphics[width=\textwidth]{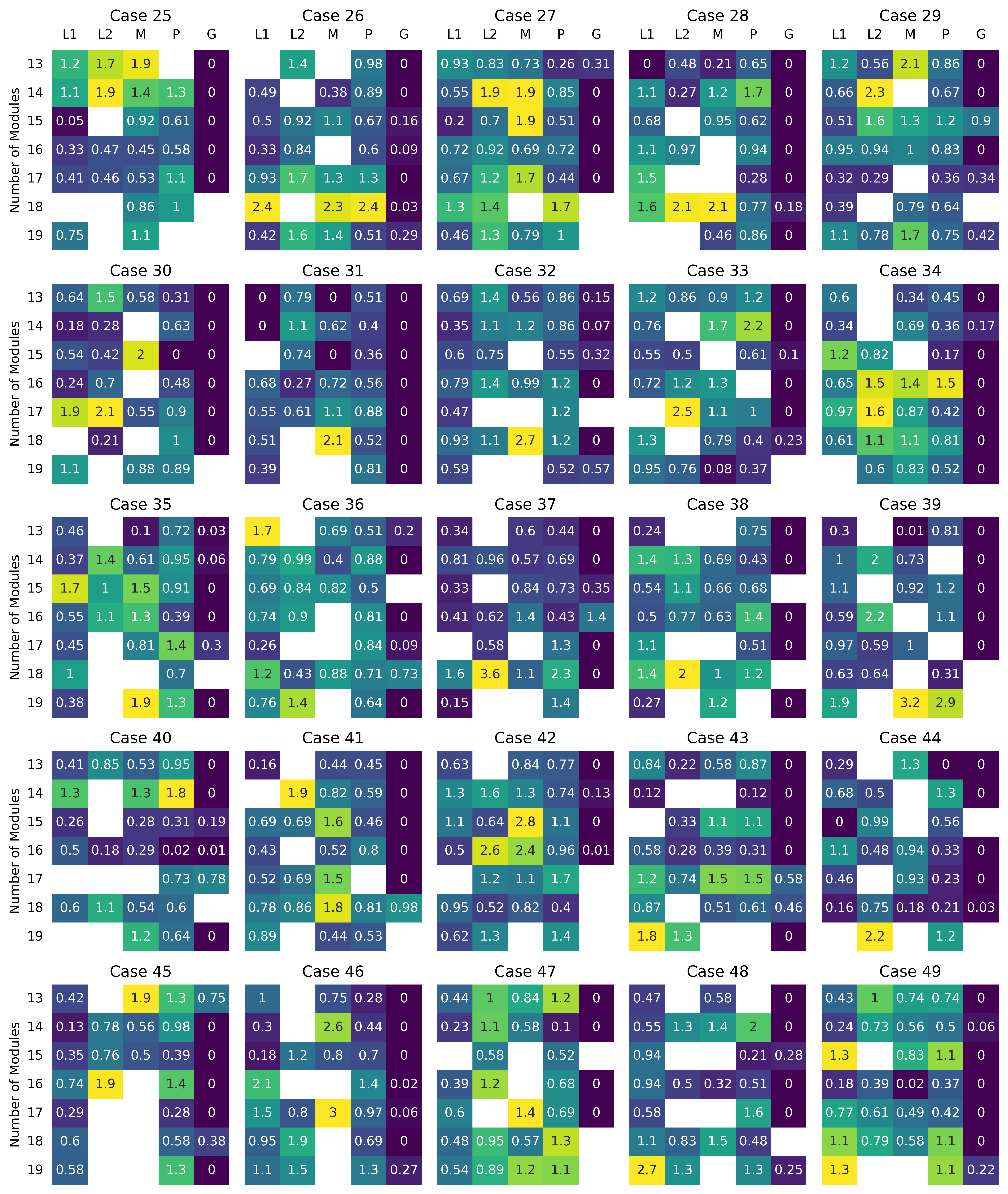}
  \caption{Original data for dead space ratios for the 16-module model (Cases 25 to 49). Brighter colors indicate lower dead space ratios. The labels "L1", "L2", "M", "P", and "G" represent "Llama3.1 (8B)", "Llama3.2 (3B)", "Mistral v0.3 (7B)", "Phi-4 (13B)", and "GPT4o-mini", respectively.}
  \Description{Plot showing dead space ratios for 16-module dataset from cases 25 to 49 across different models.}
  \label{fig:app_ori_16_1}
\end{figure}

\begin{figure}[h]
  \centering
  \includegraphics[width=\textwidth]{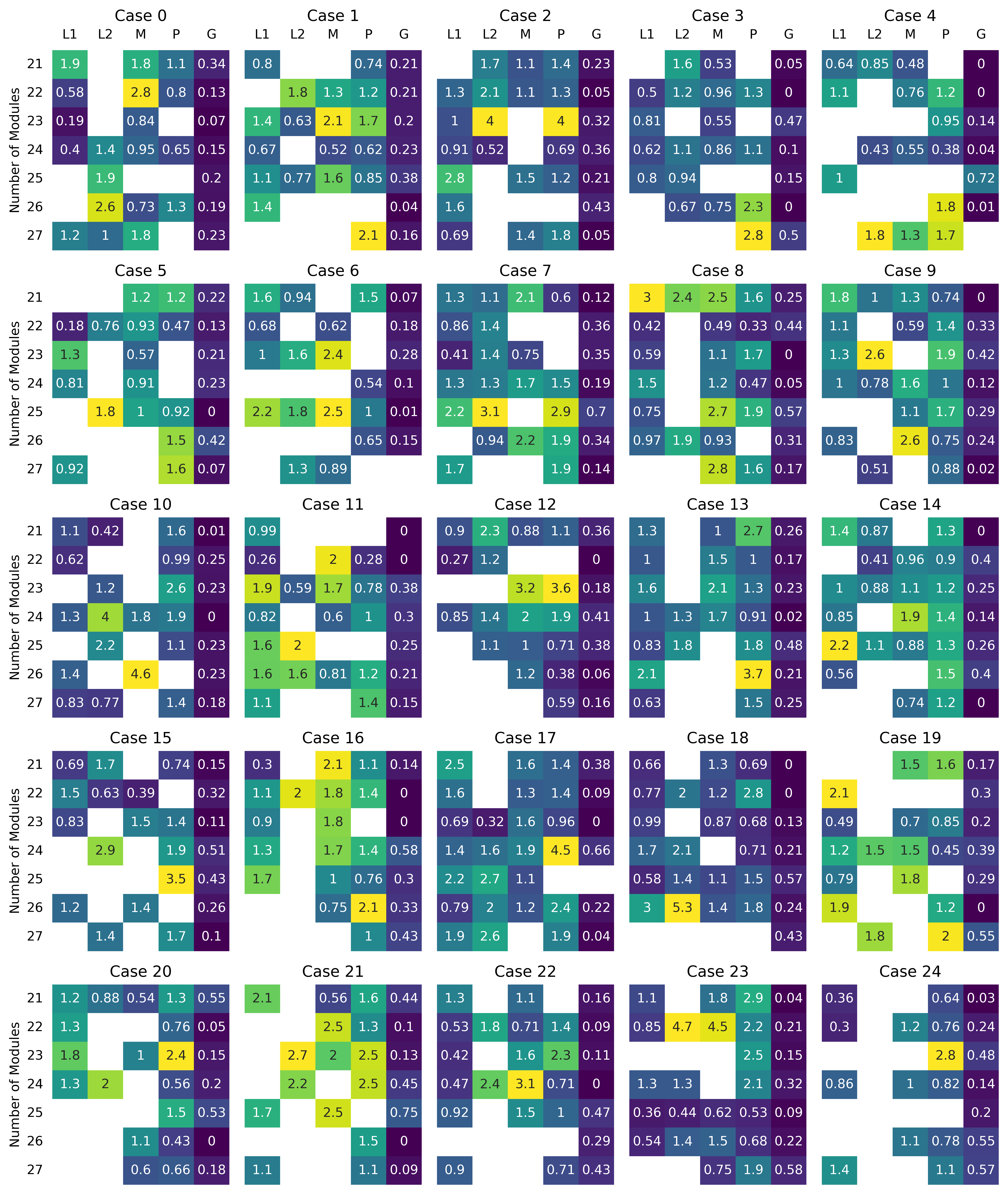}
  \caption{Original data for dead space ratios for 24-module model (Case 0 to Case 24). Brighter colors indicate lower dead space ratios. The labels "L1", "L2", "M", "P", and "G" represent "Llama3.1 (8B)", "Llama3.2 (3B)", "Mistral v0.3 (7B)", "Phi-4 (13B)", and "GPT4o-mini", respectively.}
  \Description{Plot showing dead space ratios for 24-module dataset from cases 0 to 24 across different models.}
  \label{fig:app_ori_24_0}
\end{figure}

\begin{figure}[h]
  \centering
  \includegraphics[width=\textwidth]{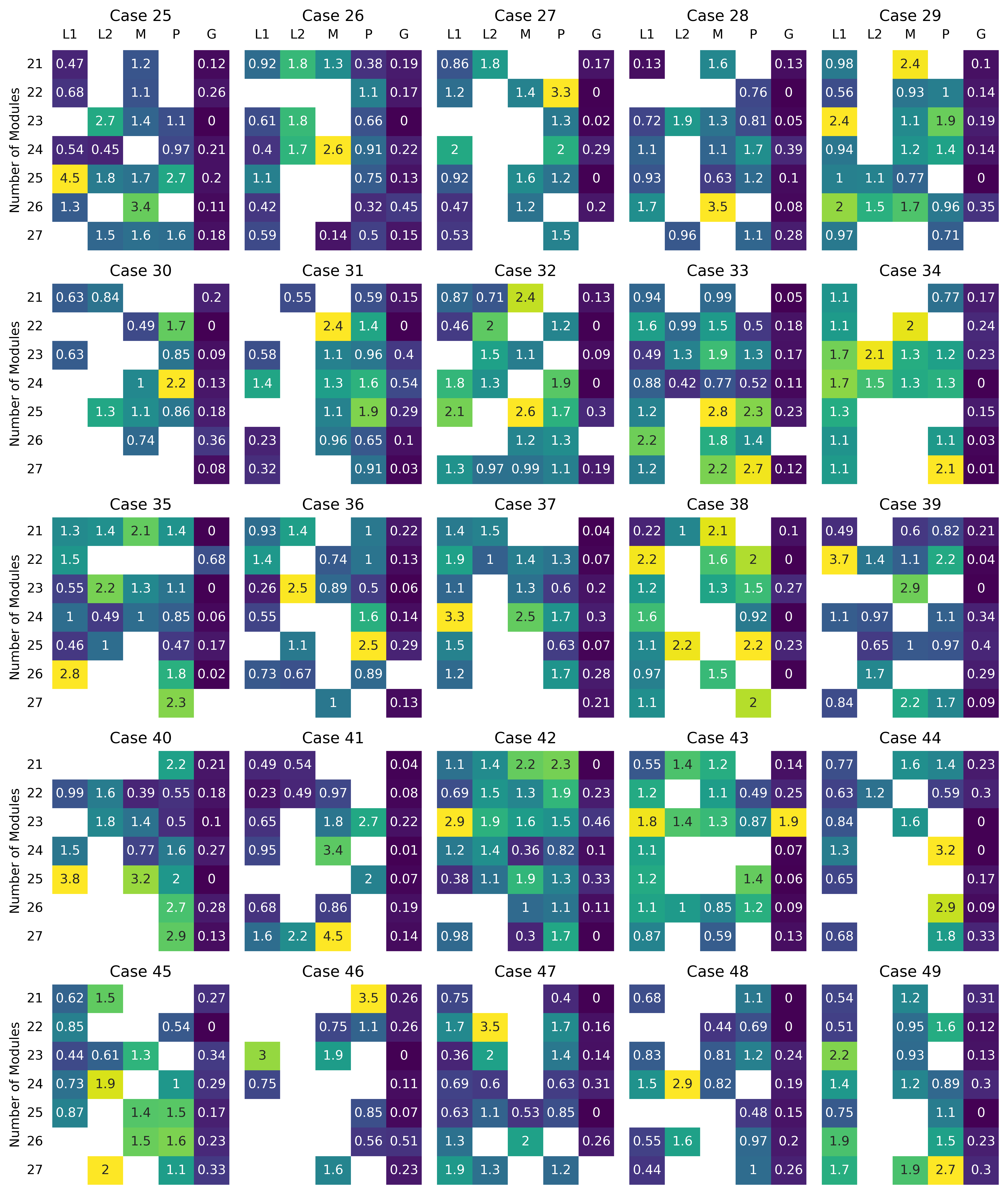}
  \caption{Original data for dead space ratios for 24-module model (Case 25 to Case 49). Brighter colors indicate lower dead space ratios. The labels "L1", "L2", "M", "P", and "G" represent "Llama3.1 (8B)", "Llama3.2 (3B)", "Mistral v0.3 (7B)", "Phi-4 (13B)", and "GPT4o-mini", respectively.}
  \Description{Plot showing dead space ratios for 24-module dataset from cases 25 to 49 across different models.}
  \label{fig:app_ori_24_1}
\end{figure}

Figure~\ref{fig:app_ori_16_0} through Figure~\ref{fig:app_ori_24_1} present the original data for the best dead space ratios obtained from the fine-tuned LLMs. Figures~\ref{fig:app_ori_16_0} and~\ref{fig:app_ori_16_1} show the dead space ratios for models fine-tuned on the 16-module dataset, while Figures~\ref{fig:app_ori_24_0} and~\ref{fig:app_ori_24_1} display the results for the 24-module dataset. The data encompasses dead space measurements across various module counts. For each module count, 50 test samples were generated, and for each test sample, the best (i.e., lowest) dead space value among 5 outputs was selected. In cases where a test sample did not produce a legal slicing tree, no dead space value is recorded (the corresponding cell remains white). The color theme used in the plots is ``viridis,'' where lighter colors indicate higher dead space ratios. By plotting the data in this manner, we can compare the dead space ratios across different models and module counts for the same test sample. In each block, each row represents a different number, and each column represents a different model. For space constraints, the x-axis labels "L1", "L2", "M", "P", and "G" represent "Llama3.1 (8B)", "Llama3.2 (3B)", "Mistral v0.3 (7B)", "Phi-4 (13B)", and "GPT4o-mini", respectively.

\clearpage

\clearpage

\clearpage

\clearpage

\clearpage

\section{Floorplanning Results}
\begin{figure}[h]
  \begin{subfigure}[b]{0.25\linewidth}
    \centering
    \footnotesize
    P\_13(243,71);P\_23(311,117); \\
    P\_26(457,69);P\_27(414,424); \\
    P\_32(243,14);P\_38(126,210); \\
    P\_40(171,35);P\_41(4,50); \\
    P\_48(394,515);P\_54(288,210); \\
    P\_55(81,117);P\_62(65,117); \\
    P\_72(457,121);P\_75(167,50); \\
    P\_84(63,515);P\_86(414,103)
    \Description{Input for 16 modules}
    \caption{Input for 16 modules}
  \end{subfigure}
  \hfill
  \centering
  \begin{subfigure}[b]{0.34\textwidth}
    \centering
    \includegraphics[width=\textwidth]{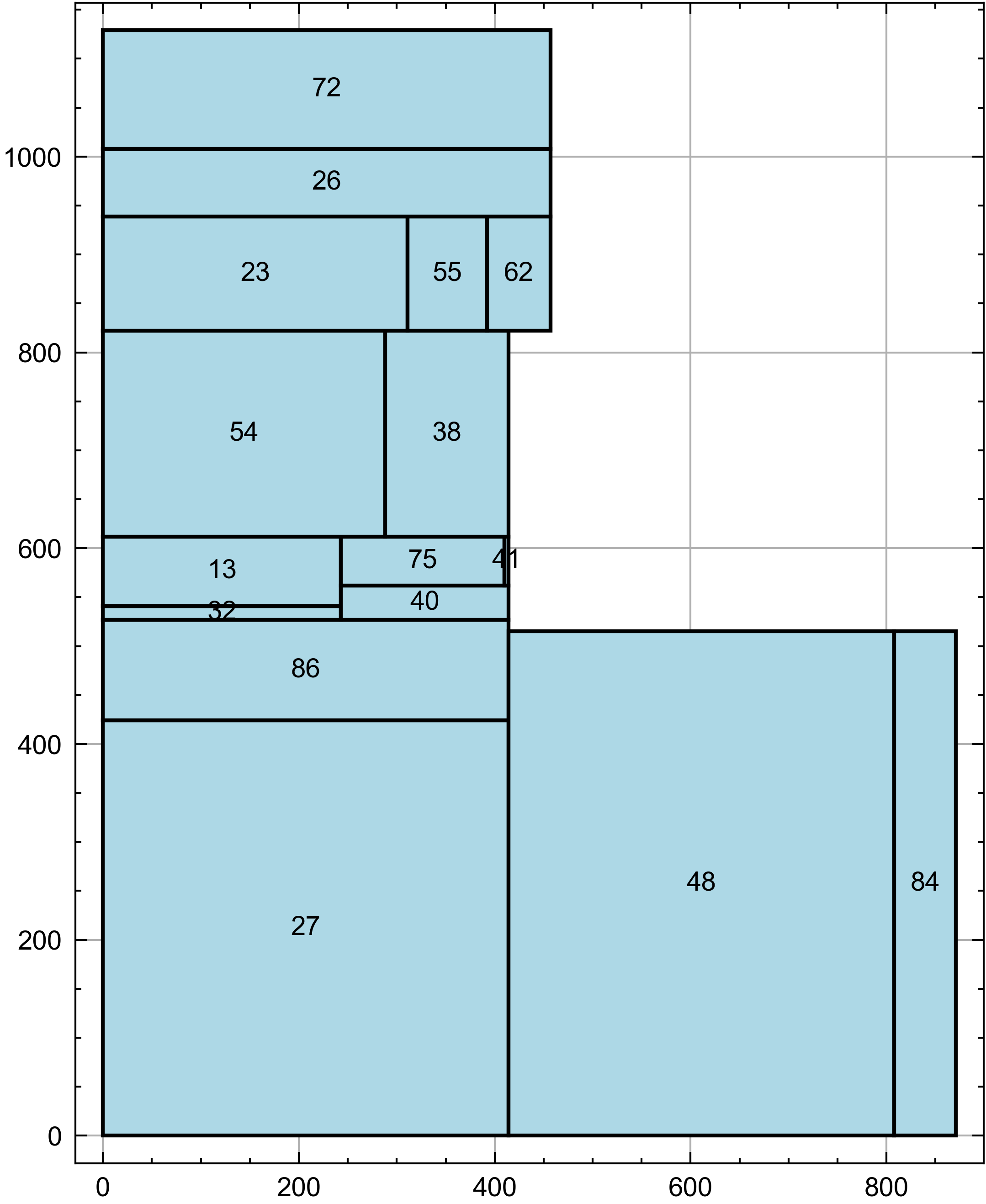}
    \caption{Non-optimal result (16 modules)}
    \Description{Floorplanning results for 16 modules generated by the fine-tuned GPT4o-mini model. The light blue rectangles represent the modules, and the label in the middle of each rectangle indicates the module identifier.}
    \label{fig:result_16_1}
  \end{subfigure}
  \hfill
  \begin{subfigure}[b]{0.34\textwidth}
    \centering
    \includegraphics[width=\textwidth]{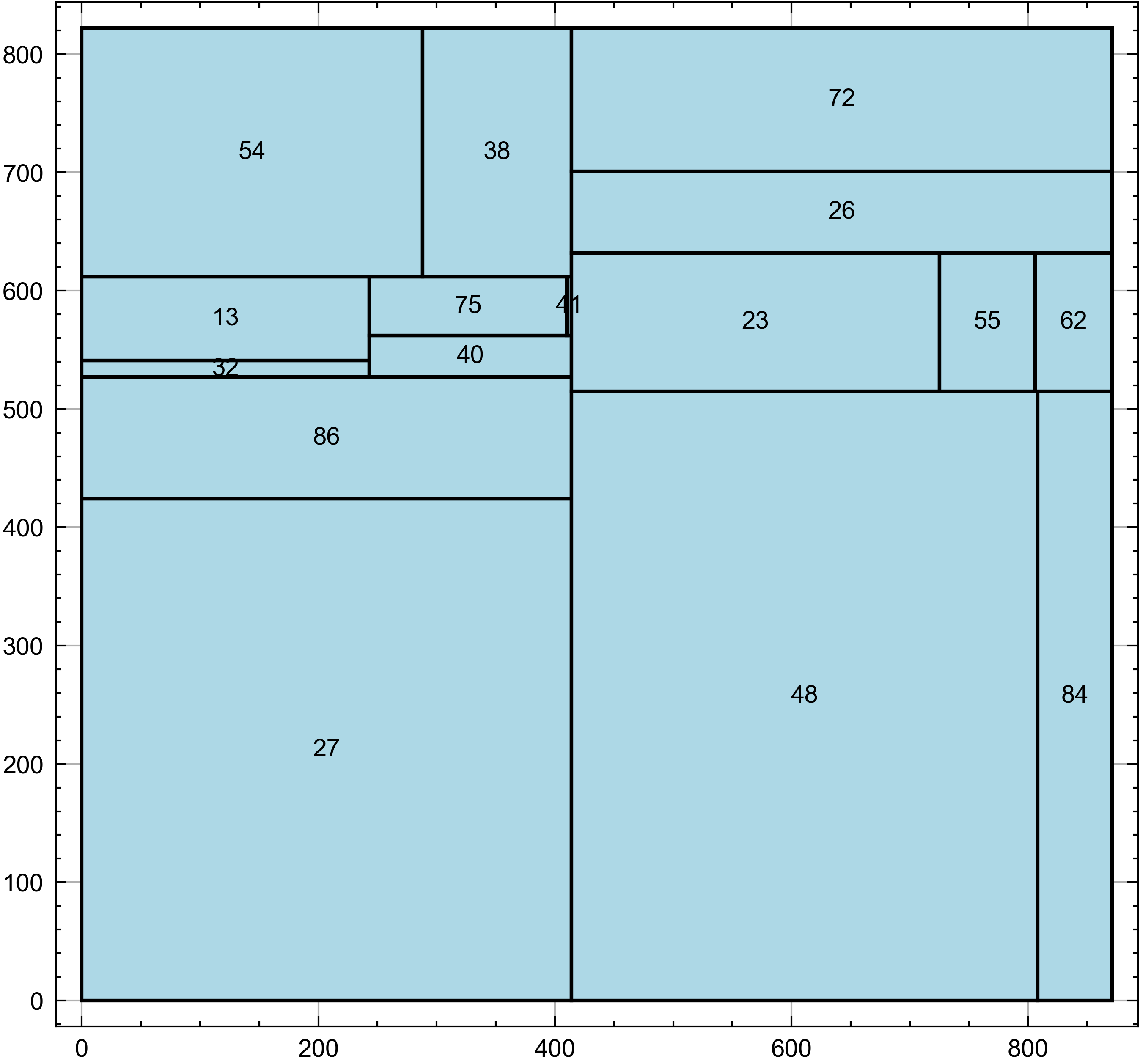}
    \caption{Optimal result (16 modules)}
    \Description{Floorplanning results for 16 modules generated by the fine-tuned GPT4o-mini model. The light blue rectangles represent the modules, and the label in the middle of each rectangle indicates the module identifier.}
    \label{fig:result_16_2}
  \end{subfigure}

  \begin{subfigure}[b]{0.25\textwidth}
    \centering
    \footnotesize
    P\_2(1187,4480); P\_7(634,4480);\\
    P\_14(2307,24120); P\_15(8834,1245);\\
    P\_20(1801,24120); P\_25(511,2333);\\
    P\_28(8834,3244); P\_34(1653,2333);\\
    P\_36(5886,2212); P\_47(3473,1059);\\
    P\_48(8834,1139); P\_49(1109,4425);\\
    P\_51(887,2333); P\_54(1100,18492);\\
    P\_55(1652,4480); P\_57(5886,6383);\\
    P\_61(422,2333); P\_66(5886,5871);\\
    P\_68(709,18492); P\_73(1139,18492);\\
    P\_77(2364,4425); P\_78(3473,9689);\\
    P\_95(3473,2134); P\_98(5886,4026)
    \caption{Input for 24 modules}
    \Description{Input for 24 modules}
  \end{subfigure}
  \hfill
  \begin{subfigure}[b]{0.34\textwidth}
    \centering
    \includegraphics[width=\textwidth]{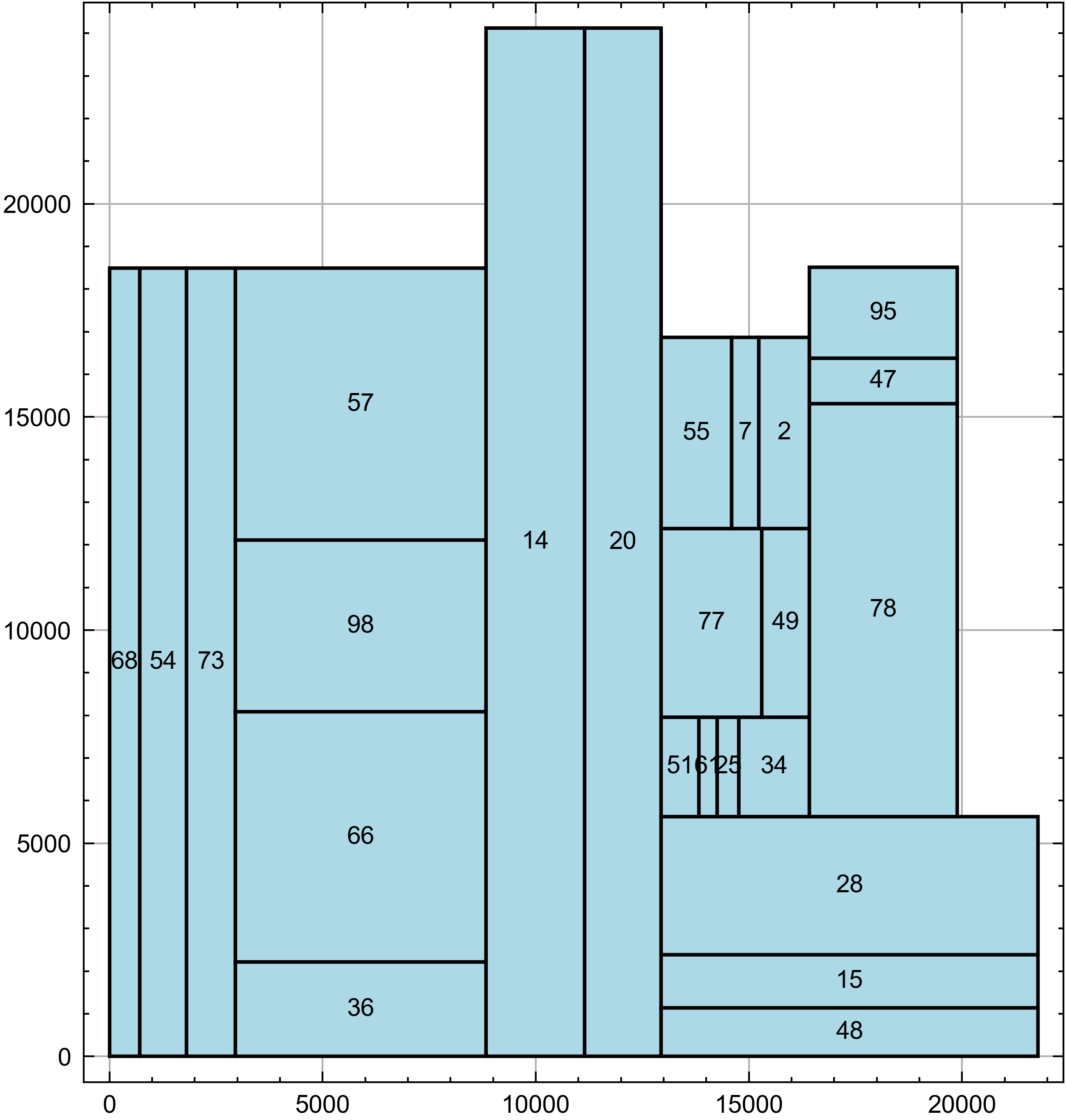}
    \caption{Non-optimal result (24 modules)}
    \Description{Floorplanning results for 24 modules generated by the fine-tuned GPT4o-mini model. The light blue rectangles represent the modules, and the label in the middle of each rectangle indicates the module identifier.}
    \label{fig:result_24_1}
  \end{subfigure}
  \hfill
  \begin{subfigure}[b]{0.34\textwidth}
    \centering
    \includegraphics[width=\textwidth]{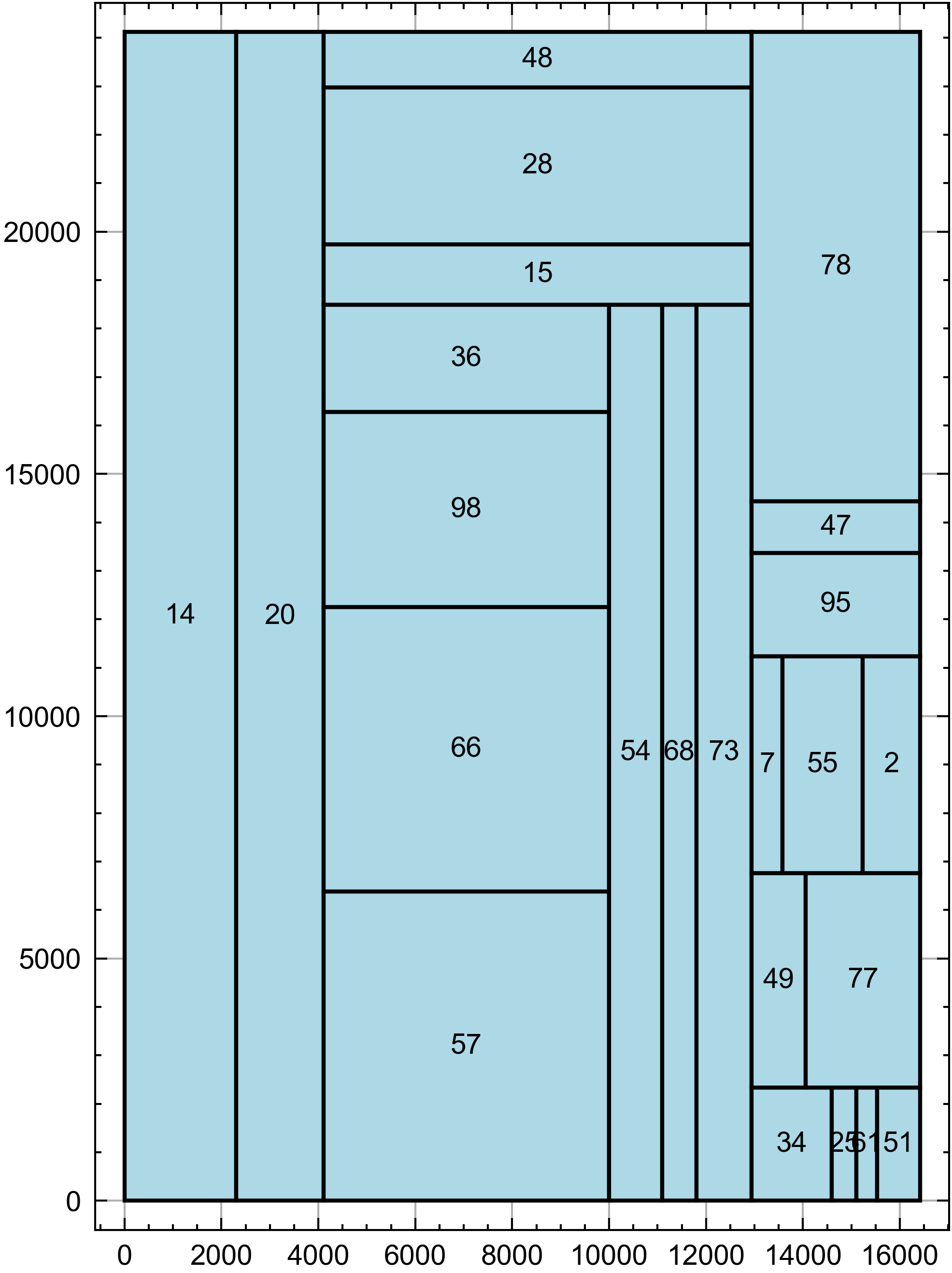}
    \caption{Optimal result (24 modules)}
    \Description{Floorplanning results for 24 modules generated by the fine-tuned GPT4o-mini model. The light blue rectangles represent the modules, and the label in the middle of each rectangle indicates the module identifier.}
    \label{fig:result_24_2}
  \end{subfigure}

  \caption{Floorplanning results for 16 and 24 modules generated by the fine-tuned GPT4o-mini model. The light blue rectangles represent the modules, and the label in the middle of each rectangle indicates the module identifier.}
  \Description{Floorplanning results for 16 and 24 modules}
  \label{fig:result}
\end{figure}

Figure~\ref{fig:result} present examples of floorplanning results generated by the fine-tuned GPT4o-mini model for the 16-module and 24-module scenarios using post-order traversal. The floorplanning problem is inherently complex; however, the model is capable of generating legal slicing trees—and, in some cases, optimal results—demonstrating the potential of LLMs in addressing such tasks. In the figures, the light blue rectangles represent the modules, and the label at the lower left corner of each rectangle indicates the module identifier.

\end{document}